\documentclass[unsortedaddress,superscriptaddress,prx,10pt,aps,twocolumn,showpacs]{revtex4-2}
\usepackage[pdftex]{graphicx}
\usepackage{graphicx}

\usepackage{microtype}
\usepackage{booktabs}
\usepackage{soul}
\usepackage[para]{threeparttable}
 \usepackage{lineno}
\usepackage[english]{babel}
\usepackage[utf8]{inputenc}
\usepackage[pdftex]{hyperref, color}

\usepackage{upgreek}
\usepackage{lmodern}
\usepackage{wrapfig}
\usepackage{dcolumn}
\usepackage{bm}

\usepackage{bbold}
\usepackage[T1]{fontenc}
\usepackage{color,bm, braket}
\usepackage{siunitx}
\usepackage[skip=0pt plus1pt, indent=20pt]{parskip}

\usepackage[table]{xcolor}
\newcolumntype{s}{ p{2.5cm}}
\arrayrulecolor[HTML]{DB5800}

\usepackage{xcolor}

\newcommand {\ce}[1]{{\color{red} #1}}

\newcommand {\mo}{MoS$_2$}

	\DeclareGraphicsExtensions{.pdf}

\begin{document}
\title{Inelastic tunneling into multipolaronic bound states in single-layer MoS\texorpdfstring{$_2$}{}}

\author{Camiel van Efferen}
\affiliation{II. Physikalisches Institut, Universit\"{a}t zu K\"{o}ln, Z\"{u}lpicher Stra\ss e 77, 50937 K\"{o}ln, Germany}
\author{Laura Pätzold}
\affiliation{I. Institut f\"ur Theoretische Physik, Universit\"at Hamburg, Notkestr.~9-11, 22607 Hamburg, Germany}
\author{Tfyeche Y. Tounsi}
\affiliation{II. Physikalisches Institut, Universit\"{a}t zu K\"{o}ln, Z\"{u}lpicher Stra\ss e 77, 50937 K\"{o}ln, Germany}
\author{Arne Schobert}
\affiliation{I. Institut f\"ur Theoretische Physik, Universit\"at Hamburg, Notkestr.~9-11, 22607 Hamburg, Germany}
\author{Michael Winter}
\affiliation{I. Institut f\"ur Theoretische Physik, Universit\"at Hamburg, Notkestr.~9-11, 22607 Hamburg, Germany}
\author{Yann in 't Veld}
\affiliation{I. Institut f\"ur Theoretische Physik, Universit\"at Hamburg, Notkestr.~9-11, 22607 Hamburg, Germany}
\author{Mark Georger}
\affiliation{II. Physikalisches Institut, Universit\"{a}t zu K\"{o}ln, Z\"{u}lpicher Stra\ss e 77, 50937 K\"{o}ln, Germany}
\author{Affan Safeer}
\affiliation{II. Physikalisches Institut, Universit\"{a}t zu K\"{o}ln, Z\"{u}lpicher Stra\ss e 77, 50937 K\"{o}ln, Germany}
\author{Christian Krämer}
\affiliation{II. Physikalisches Institut, Universit\"{a}t zu K\"{o}ln, Z\"{u}lpicher Stra\ss e 77, 50937 K\"{o}ln, Germany}
\author{Jeison Fischer}
\affiliation{II. Physikalisches Institut, Universit\"{a}t zu K\"{o}ln, Z\"{u}lpicher Stra\ss e 77, 50937 K\"{o}ln, Germany}
\author{Jan Berges}
\affiliation{U~Bremen Excellence Chair, Bremen Center for Computational Materials Science, and MAPEX Center for Materials and Processes, Universit\"at Bremen, 28359 Bremen, Germany}
\author{Thomas Michely}
\affiliation{II. Physikalisches Institut, Universit\"{a}t zu K\"{o}ln, Z\"{u}lpicher Stra\ss e 77, 50937 K\"{o}ln, Germany}
\author{Roberto Mozara}
\affiliation{I. Institut f\"ur Theoretische Physik, Universit\"at Hamburg, Notkestr.~9-11, 22607 Hamburg, Germany}
\author{Tim Wehling}
\affiliation{I. Institut f\"ur Theoretische Physik, Universit\"at Hamburg, Notkestr.~9-11, 22607 Hamburg, Germany}
\affiliation{The Hamburg Centre for Ultrafast Imaging, Luruper Chaussee 149, 22607 Hamburg, Germany}
\author{Wouter Jolie}
\email{wjolie@ph2.uni-koeln.de}
\affiliation{II. Physikalisches Institut, Universit\"{a}t zu K\"{o}ln, Z\"{u}lpicher Stra\ss e 77, 50937 K\"{o}ln, Germany}

\date{\today}
\begin{abstract}
Polarons are quasiparticles that arise from the interaction of electrons or holes with lattice vibrations. Though polarons are well-studied across multiple disciplines, experimental observations of polarons in two-dimensional crystals are sparse. We use scanning tunneling microscopy and spectroscopy to measure inelastic excitations of polaronic bound states emerging from coupling of non-polar zone-boundary phonons to Bloch electrons in n-doped metallic single-layer \mo. The latter is kept chemically pristine via contactless chemical doping. Tunneling into the vibrationally coupled polaronic states leads to a series of evenly spaced peaks in the differential conductance on either side of the Fermi level. Combining density functional (perturbation) theory with a recently developed \textit{ab initio} electron-lattice downfolding technique, we show that the energy spacing stems from the longitudinal-acoustic phonon mode that flattens at the Brillouin zone edge  and is responsible for the formation of stable multipolarons in metallic \mo.
\end{abstract}

\maketitle

\section{Introduction}
Electron-phonon coupling is known to play a do\-mi\-nant role in metallic transition metal dichalcogenides, leading to various correlated ground states such as charge density waves and superconductivity~\cite{Ugeda2015,CastroNeto2016}. Such correlation phenomena are also induced in the intrinsic semiconductor \mo~when its conduction band is filled with electrons. A charge density wave, which couples conduction electrons to unstable phonon modes, has been reported for electron-doped \mo~\cite{Ge2013, Roesner2014, BinSubhan2021, Marini2023}. Further increases of the electron density through ionic liquid gating led to the observation of a superconducting dome, under which electrons condense into Cooper pairs via electron-phonon interactions~\cite{Ge2013, Lu2015, Costanzo2016, Saito2016, Fu2017, Costanzo2018}. Finally, at even higher charge carrier concentrations the entire crystal lattice becomes unstable, leading to an electronically driven structural phase transition from the 1H to the 1T structure~\cite{Zhuang2017}.

Polarons represent local manifestations of electron-phonon coupling~\cite{Emin2013}. These many-body quasiparticles form when charge carriers interact strongly with the ionic lattice, creating a localized state in absence of a defect~\cite{Franchini2021}. A polaron is generally able to move through the lattice, though it drags a deformation of the lattice with it, enhancing its effective mass. This mass-enhancement has been observed in bulk \mo~doped with rubidium using angle-resolved photoemission spectroscopy (ARPES) and was attributed to the formation of multiple Holstein polaron bands~\cite{Kang2018}. A follow-up \textit{ab initio} study could however reproduce the experimental spectral function without small polaron formation, but using the Fan-Migdal perturbative self-energy. This approach, which is geared towards the weak coupling limit, suggested that the mass renormalization stems from interactions with two different M-point phonons~\cite{Garcia-Goiricelaya2019}. The phonons couple electronic states close to the conduction band minima at K and K$'$ with states near Q$'$ and Q. These local minima of the conduction band are only $\approx100$~meV higher than the global conduction band minimum and also partly filled in the ARPES experiment~\cite{Kang2018}.

The proposed importance of the local minimum at Q for polaron formation in bulk \mo{} poses the question whether polarons can actually form in single-layer (SL) \mo, since the energy difference between K and Q is much larger in n-doped SLs~\cite{VanEfferen2022, Caruso2021, Khestanova2023}. In addition, a recent theoretical study postulated that stable polarons do not form in SL \mo~\cite{Sio2023}, although only monopolarons were considered in these calculations. The physics of multipolarons, i.e. bound states formed by two or more electrons coupled to phonons, is in general much less studied, though it is postulated to become dominant in cases of suppressed Coulomb potentials and sufficiently large electron-phonon coupling \cite{Verbist1992,Frank2010}. While signatures of plasmonic polarons have been observed in highly defective SLs of \mo~\cite{Caruso2021}, the observation of polarons in defect-free SL \mo~remains elusive.

  \begin{figure*}
	\centering
		\includegraphics[width=0.95\textwidth]{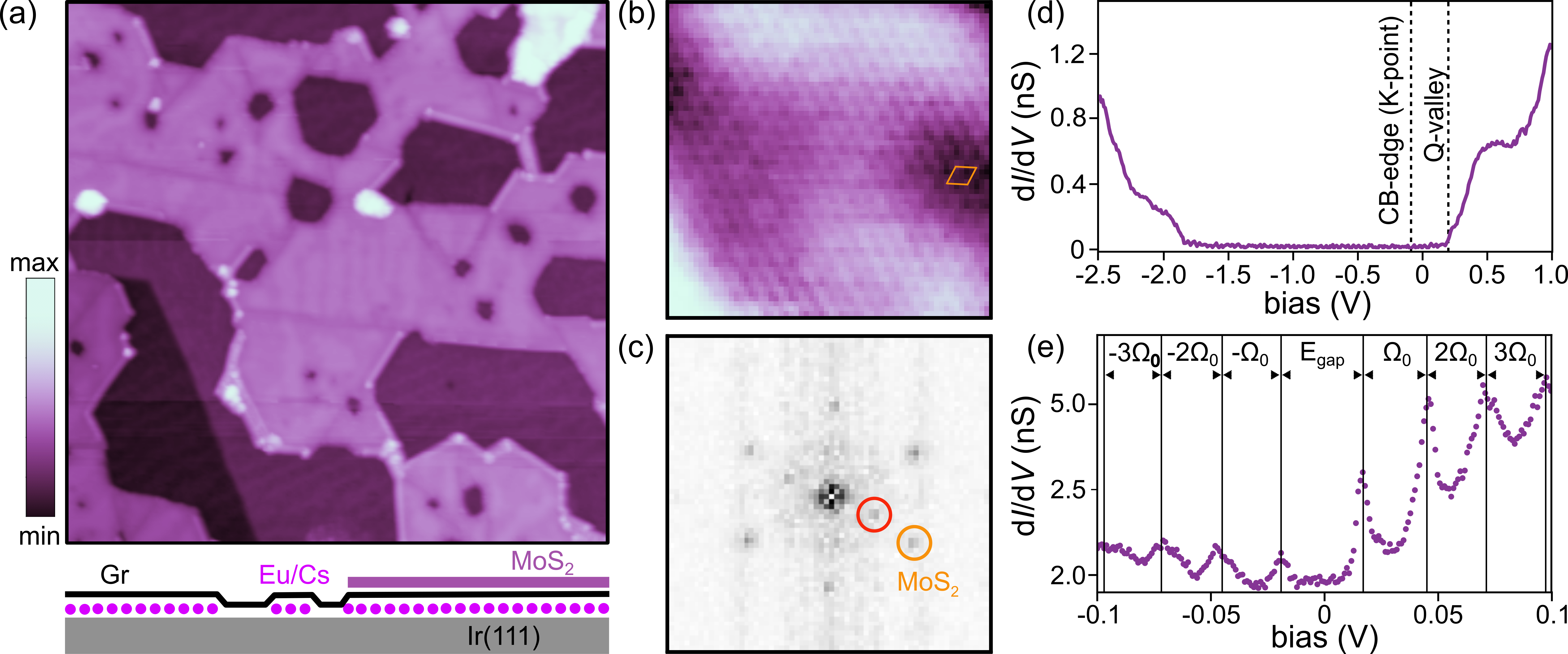}
			\caption{Electronic structure of SL \mo. (a)~Large-scale STM image of SL \mo/Gr/Eu/Ir(111). A step edge in the Ir substrate can be seen in the bottom left. The Eu intercalation has slight density modulations which lead to the stripes visible through \mo{} and Gr. (b) Atomic resolution image of \mo, revealing a local superstructure. The \mo~unit cell is indicated. (c) Fourier transformation of (b), showing the six spots of the reciprocal \mo~lattice together with additional features at wave vectors close to half the size of the reciprocal lattice. Both features are marked with circles. (d)~Large-bias d$I$/d$V$ spectrum of single-layer \mo{} on Gr/Eu/Ir(111). The conduction band edge at the K-point of the Brillouin zone and the onset of the valley at 0.2~eV near the Q-point are indicated. (e)~Low-bias STM d$I$/d$V$ spectrum, revealing peak-dip features and a gap at the Fermi level, characteristic for metallic SL \mo. 
			STM/STS parameters: (a)~\SI[parse-numbers=false]{100 \times 100}{\nano\meter\squared}, $V_\text{set} = \SI{2.5}{\V}$, $I_\text{set} = \SI{50}{\pA}$;  
            (b)~\SI[parse-numbers=false]{5 \times 5}{\nano\meter\squared}, $V_\text{set} = \SI{0.1}{\V}$, $I_\text{set} = \SI{500}{\pA}$;
						(d)~$V_\text{set} = \SI{1}{\V}$, $I_\text{set} = \SI{500}{\pA}$, $V_\text{mod} = \SI{15}{\mV}$; (e)~$V_\text{set} = \SI{0.1}{\V}$, $I_\text{set} = \SI{500}{\pA}$, $V_\text{mod} = \SI{1}{\mV}$, smoothed using the Savitzky-Golay method. For all spectra $f_\text{mod} = \SI{877}{\Hz}$.
}
 \label{fig:Fig1}
\end{figure*}

Here, we investigate the role of electron-phonon coup\-ling and its link to multipolarons in SL \mo{} on a graphene (Gr) on Ir(111) substrate using scanning tunneling microscopy (STM), scanning tunneling spectroscopy (STS), first-principles density functional [perturbation] theory (DF[P]T) calculations and a newly developed multi-scale technique based on downfolding for electron-phonon coupled systems \cite{Schobert2024}. Intercalating Cs and Eu between Gr and Ir(111) shifts the conduction band of \mo{} below the Fermi energy $E_\text{F}$, thus inducing an insulator-metal transition~\cite{VanEfferen2022}. This enables us to study the intrinsic properties of metallic SL \mo, which rests on a weakly interacting substrate (Gr).

\section{Results}
\subsection{Scanning tunneling spectroscopy}
An STM topograph of metallic SL \mo{} is shown in Fig.~\ref{fig:Fig1}~(a), displaying the typical morphology. The \mo{} islands have merged during growth\ce{,} forming extended networks, with defect-free areas ranging in size from \SIrange{50}{700}{\nm^2}. These areas are separated by grain boundaries, which are imaged as dark stripes when tunneling in the conduction band of \mo. In addition, faint stripes can be recognized on and next to \mo, which stems from minor density variations in the Eu intercalation layer below graphene, as discussed previously~\cite{VanEfferen2022}.

Our first experimental observation related to electron-phonon-physics is a weak local superstructure appearing in atomically resolved STM images, presented in Fig.~\ref{fig:Fig1}~(b). While the superstructure is locally visible in real space, there is no long-range coherence of the charge density, ruling out the presence of a charge density wave. The corresponding Fourier transformation in Fig.~\ref{fig:Fig1}~(c) reveals a broad feature close to the ($2\times2$) spot in reciprocal space. This intensity has no structural origin, as intercalated Cs forms a disordered structure after annealing, while intercalated Eu forms a $c(4\times2)$ superstructure with respect to graphene. These intercalation patterns are not visible on \mo, as both Cs and Eu layers are intercalated below graphene~\cite{VanEfferen2022}. As we will show below, multipolarons in \mo~ are expected to generate local lattice distortions with the same dominant wave vector as the local ($2\times2$) superstructure observed experimentally.

A d$I$/d$V$ spectrum showing the band gap of \mo{} is depicted in Fig.~\ref{fig:Fig1}~(d). The spectrum shows an apparent onset of the conduction band above the Fermi level.  However, since the conduction band minimum of doped SL \mo{} lies at the K-point~\cite{Ge2013, Khestanova2023}, the high parallel momentum $k_{||}$ of the quasiparticles leads to a strongly reduced tunneling probability, unless the tip-sample distance is severely reduced~\cite{Zhang2015, Murray2019}. Indeed, the low-bias d$I$/d$V$ spectrum in Fig.~\ref{fig:Fig1}~(e) reveals a finite spectral intensity in the occupied states, separated from the unoccupied states by a gap at $E_\text{F}$. Comparable spectra are measured on Cs-intercalated samples. On account of the n-doping of the Gr substrate layer by the Eu atoms, the \mo{} conduction band edge is shifted to $-90$~meV below the Fermi energy, while the local minima at Q lies above the Fermi energy~\cite{VanEfferen2022}.

\begin{figure*}
	\centering
		\includegraphics[width=0.8\textwidth]{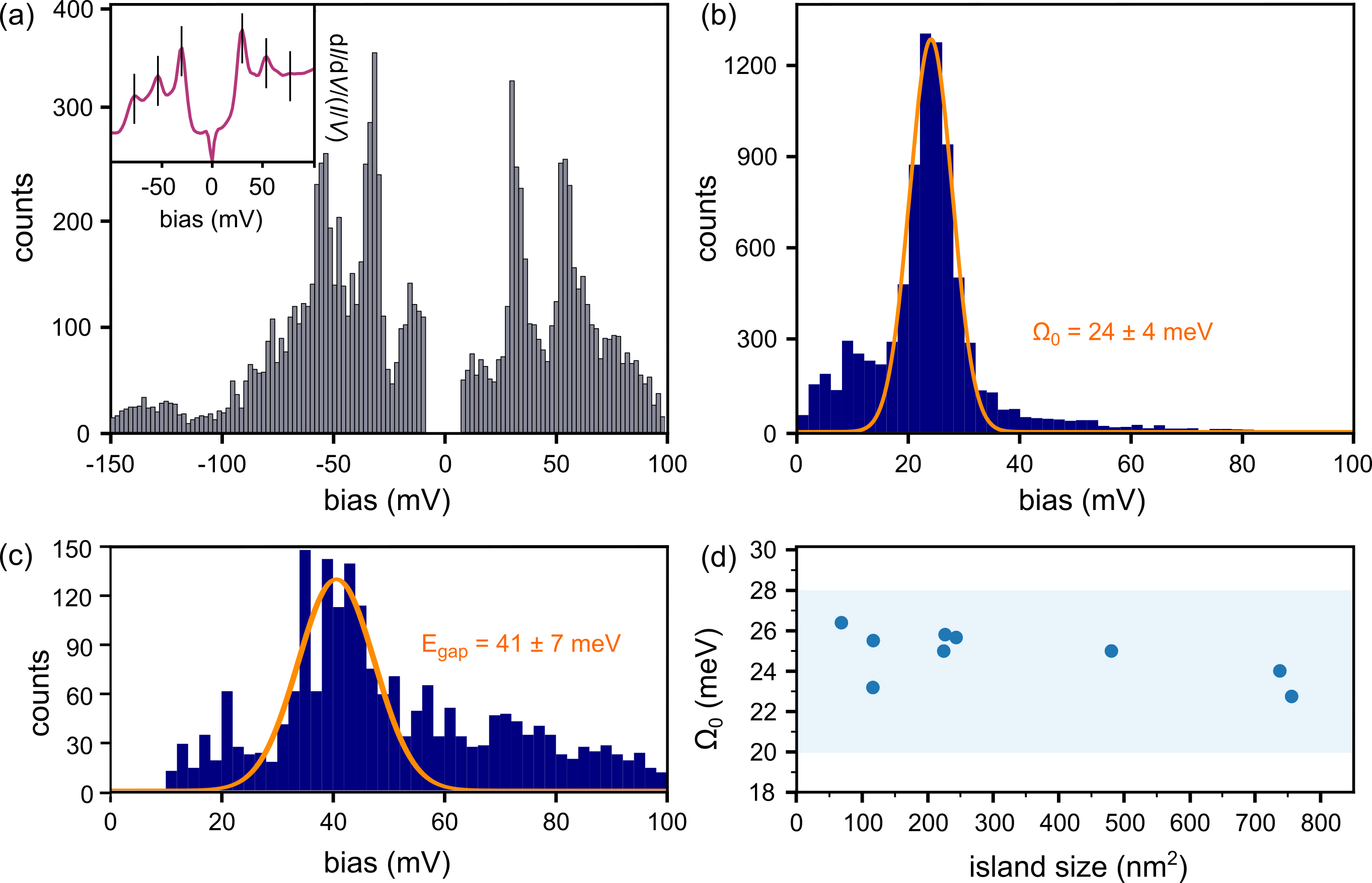}
			\caption{Statistical analysis of peak-dip features in d$I$/d$V$ spectra. (a)-(b)~Histograms of 2935 d$I$/d$V$ spectra acquired on different locations, at different temperatures (\SIrange{0.4}{6.5}{\K}) and on \mo{} areas of different sizes, using different samples (multiple \mo/Gr/Eu/Ir(111) samples and one \mo/Gr/Cs/Ir(111) sample) and tips (Au and PtIr).  In (a) the peak locations are plotted. To analyze the data, the d$I$/d$V$ spectra were smoothed and normalized. The inset in (a) shows a d$I$/d$V$ spectrum after this procedure, with the locations of peaks found by a peak finding algorithm marked. In (b) the energetic distances between neighbouring peaks (excepting the peaks closest to the Fermi level) are plotted. A Gaussian fit to the peak yields $\Omega_0 = \SI{24 \pm 4}{\meV}$. (c)~Histogram of the distribution of the energetic distance between the first peaks on either side of the Fermi level. A Gaussian fit to the peak yields $E_\text{gap} = \SI{41 \pm 7}{\meV}$. (d)~Average distance between neighbouring peaks $\Omega_0$ for spectra taken on \mo{} islands of different sizes. The island size is the coherent island area taking into account edges, grain boundaries and constrictions smaller than 10~nm. The shaded area represents $\Omega_0 = \SI{24 \pm 4}{\meV}$. 
			}
 \label{fig:Fig2}
\end{figure*}

A particular aspect of the spectrum in Fig.~\ref{fig:Fig1}~(e) is the presence of peak-dip features close to $E_\text{F}$. 
These are seen to manifest as a series of evenly spaced peaks, present below and above $E_\text{F}$. The peaks are spaced by a fixed energy $\Omega_0$, with a larger gap of width $E_\text{gap}$ separating the states around $E_\text{F}$. The peak-dip features are absent in semiconducting \mo, see Appendix~\ref{absence}.

To better understand the behavior of the peaks, we have analysed approximately 3000 d$I$/d$V$ spectra taken on islands of different size and on two different metallic \mo{} samples -- with either Eu or Cs intercalated below Gr. We included spectra taken in a temperature range between 0.4 and 6.5 K. In Fig.~\ref{fig:Fig2}~(a), the resulting peak positions are plotted in a histogram. The peaks have a clear tendency to be located at symmetric positions with respect to the Fermi energy (0~V), such as $\pm 32$~mV and $\pm55$~mV. We further analyze the single spectra and find that the spacing between peaks located above and below the Fermi energy has a Gaussian distribution, with a mean value and standard deviation of $\Omega_0 = \SI{24 \pm 4}{\meV}$, see Fig.~\ref{fig:Fig2}~(b). We therefore interpret all these peaks as harmonics of a bosonic mode $\Omega_0$ instead of independent excitations. When we check for the size of $E_\text{gap}$ across all investigated spectra, see Fig.~\ref{fig:Fig2}~(c), we find considerably more fluctuations than for the peak spacing. The distribution does have a marked peak, which fitted with a Gaussian gives $E_\text{gap} = \SI{41 \pm 7}{\meV}$.

We have checked in Fig.~\ref{fig:Fig2}~(d) that there is no dependence of $\Omega_0$ on the size of the coherent island. Since the grain boundaries that divide the different regions of the \mo{} islands are charged relative to their surroundings, band bending effects take place perpendicular to them, isolating the different parts of the structure from one another~\cite{Murray2020}. The sample is thus composed of many areas of differing size, all of which show the same peak spacing $\Omega_0$, ensuring that the peak-dip features are not related to area-dependent effects like confinement. The peaks are also independent of quasiparticle scattering effects within the metallic \mo{} band, see Appendix~\ref{QPI}.

\begin{figure*}[t]
	\centering
		\includegraphics[width=1\textwidth]{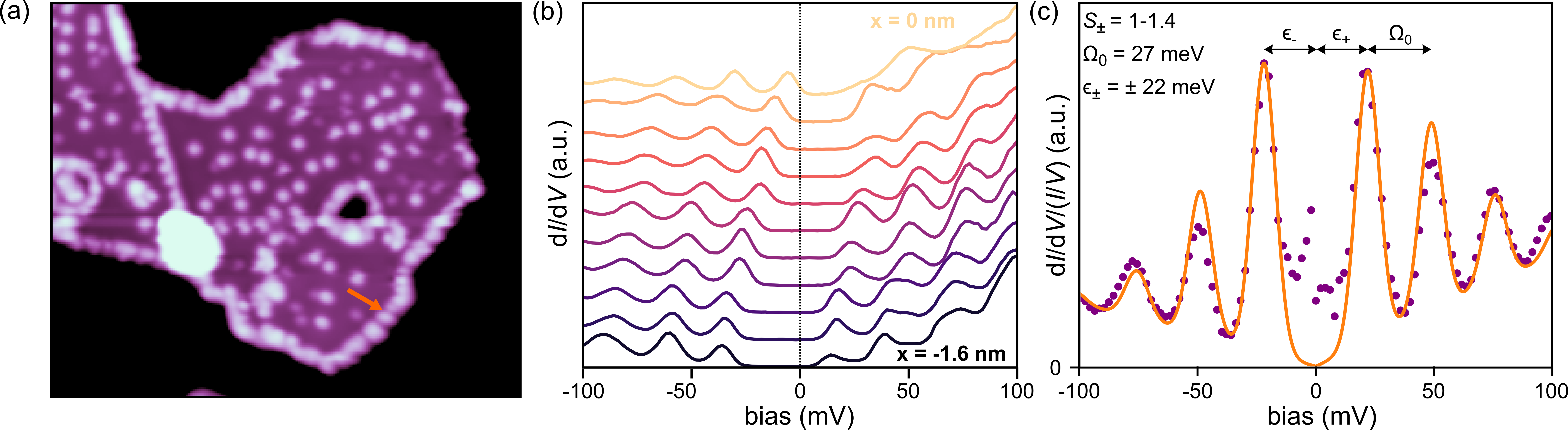}
			\caption{Tunneling through localized polaronic states. (a)~STM image of a SL \mo{} island on Gr/Cs/Ir(111). Adsorbed Cs atoms are visible on the island, grain boundaries and edges as bright dots. (b)~d$I$/d$V$ spectra taken along the orange arrow in (a), with $x = \SI{0}{\nm}$ corresponding to the spectrum closest to the edge. Spectra are offset for clarity. (c)~Low-bias d$I$/d$V$/($I$/$V$) spectrum with peaks above and below the Fermi level (purple dots). A fit based on Ref.~\citenum{Cochrane2021} is shown as an orange line. Fitting parameters: the Huang-Rhys factor $S_- =1$ ($S_+ =1.4$) for the resonances in the occupied (unoccupied) states, the phonon energy $\Omega = \SI{27}{\meV}$, the location of the polaronic state $\varepsilon = \SI{\pm22}{\meV}$, the full width at half maximum of the states $\Gamma = \SI{7}{\meV}$, $l\leq10$. A linear offset is included to account for (inelastic) tunneling processes that involve the \mo{} band.
			STM/STS parameters: (a)~\SI[parse-numbers=false]{40 \times 33}{\nano\meter\squared}, $V_\text{set} = \SI{1.0}{\V}$, $I_\text{set} = \SI{100}{\pA}$; (b)-(c)~$V_\text{set} = \SI{100}{\mV}$, $I_\text{set} = \SI{1.1}{\nA}$, $V_\text{mod} = \SI{2}{\mV}$, $f_\text{mod} = \SI{877}{\Hz}$.
			}
 \label{fig:Fig3}
\end{figure*}

\subsection{Effect of charge inhomogeneities}

While the relative energy spacing of the peaks is locally reproducible, their absolute energetic positions vary depending on the location of data acquisition. These variations are especially pronounced close to charge inhomogeneities. To visualize this finding, we measure d$I$/d$V$ spectra on metallic \mo{} with additional Cs atoms adsorbed on top, shown in Fig.~\ref{fig:Fig3}~(a). The Cs adatoms are imaged with the STM as bright spots, which are preferentially decorating the island edges and grain boundaries. Due to the small electronegativity of Cs, the adatoms donate electrons to their environment, n-doped \mo. Cs is also intercalated between graphene and Ir(111) to further ensure that \mo{} is metallic, see Appendix~\ref{doping}. Taking a series of d$I$/d$V$ spectra along the line drawn in Fig.~\ref{fig:Fig3}~(a), we can track the behavior of the peak-dip features as the tip moves towards the island edge. The spectra are plotted in Fig.~\ref{fig:Fig3}~(b). The peaks are seen to bend collectively upwards near the edge of the island in a range of \SI{1.6}{\nm}, while not crossing the Fermi energy and thus retaining a finite gap $E_\text{gap}$.

The behavior in close vicinity to randomly placed Cs atoms on the \mo~islands is found to be more complex, see Appendix \ref{Cslinescan}, likely due to the inhomogeneous electrostatic environment there.

\subsection{Modelling inelastic tunneling processes}
Given the experimental evidence, we can conclude that tunneling in metallic SL \mo{} is strongly affected by the existence of a bosonic mode of energy $\Omega_0 = \SI{24 \pm 4}{\milli\eV}$, leading to pronounced satellite peaks in tunneling spectra. Since SL \mo{} hosts multiple flat phonon bands near the edges of its Brillouin zone, with energies of \SIrange{20}{30}{\meV}~\cite{Tornatzky2019}, the bosonic mode $\Omega_0$ is likely a phonon. While the satellites are pinned to the peaks $\varepsilon_{\pm}$~defining $E_\text{gap}$, the latter is found to continuously shift up or down in energy relative to the Fermi energy, depending on the electrostatic environment. These observations point towards an inelastic tunneling process for the satellites, while $E_\text{gap}$ seems related to an intrinsic property of metallic \mo.

Inelastic tunneling processes can lead to characteristic satellites in d$I$/d$V$ when electrons tunneling through a localized level experience strong electron-phonon coup\-ling. This has been established for STM junctions in 1D and 0D systems~\cite{Lundin2002, Galperin2004, Skorobagatko2012}, in particular for molecules on surfaces~\cite{Stipe1998,Franke2012, Krane2018} and in tunneling devices based on semiconductor quantum wells~\cite{Zou1992}. In these systems, a series of resonant peaks spaced by the energy of the vibronic mode are observed in conductance spectra, in close resemblance to our observations. To obtain information related to these inelastic tunneling processes, we fit our peak-dip features to the many-body spectral function of a discrete electronic state with energy $\varepsilon_{\pm}$ coupled to a bosonic phonon mode $\Omega_0$~\cite{Wingreen1988,Cochrane2021}:

\begin{equation}
{A}(\omega )=2\pi \mathop{\sum }\limits_{{l=1}}^{\infty }\left[\left({e}^{-{S_{\pm}}}\frac{S_{\pm}^{l}}{l!}\right)\delta \left(\hbar \omega -{\varepsilon_{\pm}}\mp\Omega_0{l}\right)\right].
\label{Eq:1}
\end{equation}

From the fit we obtain the Huang-Rhys factor $S_{\pm}$, which is related to the electron-phonon coupling strength $g= \sqrt{S_{\pm}}\,\Omega_0$ and $l$ being an integer. We obtain Huang-Rhys factors of $S_- =1$ and $S_+ =1.4$ for the resonances in the occupied and unoccupied states of the spectrum in Fig.~\ref{fig:Fig3}~(c), respectively. We find a range of $S_{\pm} = 1$--$3$, using several representative spectra, see Appendix~\ref{Sfitting}. Though the model is able to adequately describe the experimental spectra and points to strong electron-phonon coupling involving a well-defined mode $\Omega_0$, it does not provide an explanation for the origin of the energy states $\varepsilon_{\pm}$ around $E_\text{F}$, nor why a single phonon energy is dominating in the experimental spectrum.

\begin{figure*}
	\centering
		\includegraphics[width=1.0\textwidth]{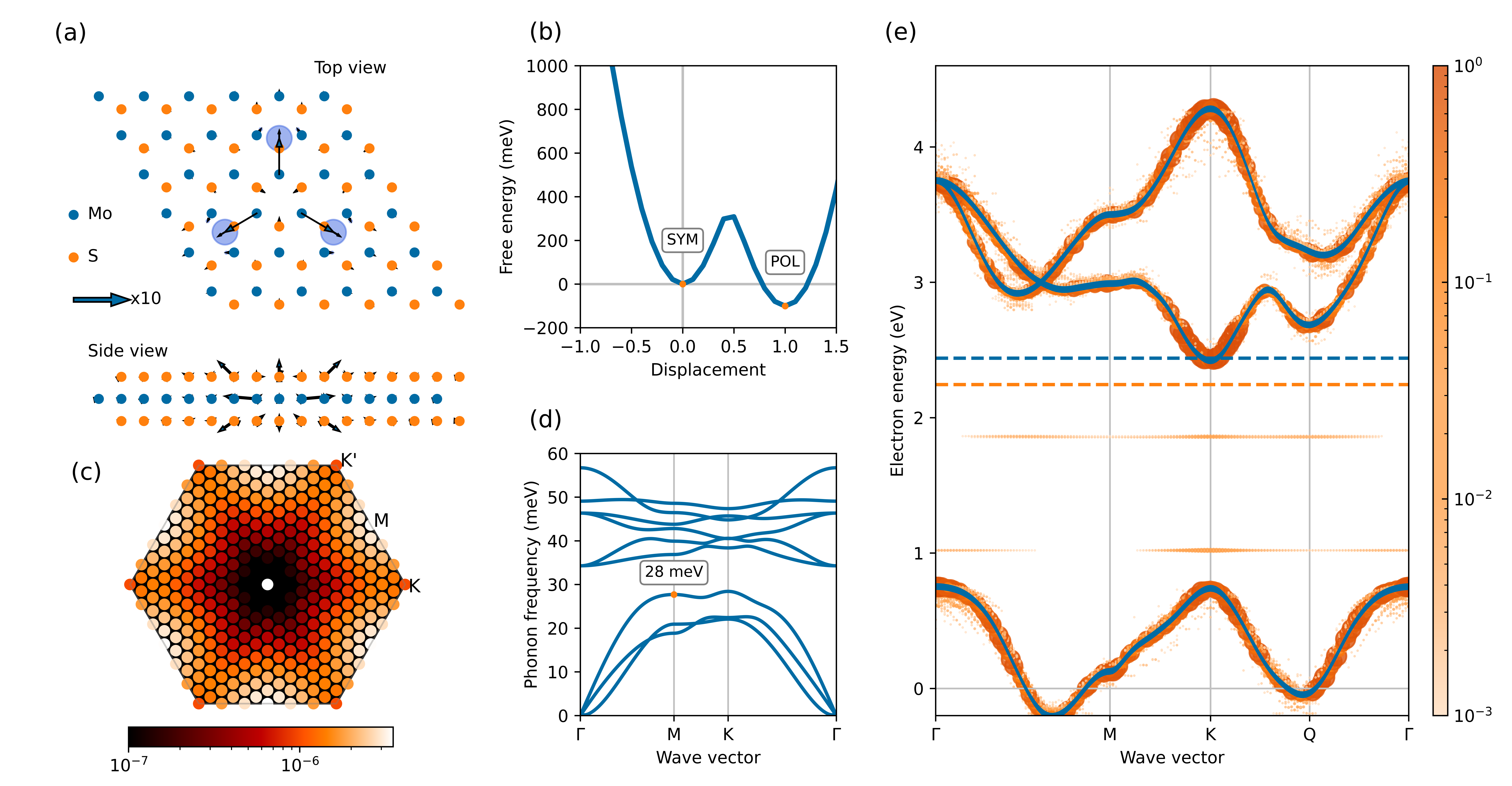}
			\caption{Multipolaronic deformations in metallic SL \mo.~(a) Zoom into relaxed crystal structure based on \textit{ab initio} model calculations on an $18 \times 18$ supercell with a lattice constant of $\SI{3.19}{\AA}$ at an electronic doping of $\SI{0.0124}{e^{-}}$ per formula unit of \mo{}. The doping is slightly increased compared to the one used in DFT+DFPT calculations ($\SI{0.01}{e^{-}}$/\mo{}) and results in four additional electrons per supercell. The displacements visualized by arrows are rescaled by a factor of 10 for better visibility. Blue circles indicate the localization of the additional electrons. (b)~Total energy as function of displacement for the polaronic deformation shown in (a). The relaxation started from the symmetric phase (SYM) and shows a minimum at the displacement corresponding to the multipolaronic deformation (POL). The resulting total energy gain of $ \approx100$~meV equals the binding energy of the multipolaronic distortion. (c)~Structure factor $S(\mathbf q)$ for the relaxed geometry seen in (a). The high intensity around the M-points of the phononic Brillouin zone shows that the modes located there contribute the most to the polaronic deformation. (d)~Phonon bands obtained via DFPT. The LA phonon mode that contributes the most to the polaronic deformation is marked by its energy at the M-point. The phonon bands are obtained at an electronic doping of $\SI{0.01}{e^{-}}$/MoS$_2$. (e)~Band structure of low-energy subspace as obtained by DFT (blue) and after the \textit{ab initio} model relaxation (orange). The weights of the unfolded bands correspond to the logarithmic colorbar and the size of the markers. The relaxed band structure has been unfolded from the super- to the unit cell, revealing the opening of small gaps due to the formation of dispersionless states associated with the symmetry breaking due to the polaronic deformation, which also lowers the Fermi energy (dashed lines).
}
 \label{fig:Fig4}
\end{figure*}

\subsection{Polaronic deformations in metallic \mo}
 To unveil the origin of the pronounced peak-dip features observed in STS, we performed a theoretical analysis combining density functional theory (DFT) and density functional perturbation theory (DFPT) with a recently developed \textit{ab initio} electron-lattice downfolding technique which facilitates relaxations of large supercells~\cite{Schobert2024} (see Appendix~\ref{sec:calculation_details} below for more information) and compared to the scenarios put forward in Refs. \cite{Kang2018,Garcia-Goiricelaya2019}.

The ARPES study of Ref.~\citenum{Kang2018} revealed dip-peak features in the photoemission spectral function, that were explained in Ref.~\citenum{Garcia-Goiricelaya2019} as Fan-Migdal self-energy effects stemming from scattering of electrons between K and Q$'$ as well as K$'$ and Q valleys. Following this idea, we calculate the Fan-Migdal self-energy and show the resulting electronic spectral function for the system at hand in Appendix~\ref{spectralfunction}. For a Fermi level such that K- and Q-valleys of the SL \mo{} conduction band are occupied, we find kinks and gaps in the spectral function, very similar to Ref.~\citenum{Garcia-Goiricelaya2019}. However, our STS experiments (see Fig.~\ref{fig:Fig1}) locate the Q-valley \SI{\sim 150}{\milli\eV} above the Fermi level. In this situation, the gaps in the electronic spectral function disappear, see Appendix~\ref{spectralfunction}. Hence, the peak-dip features seen here in STS (Figs.~\ref{fig:Fig1}~(e), \ref{fig:Fig2} and \ref{fig:Fig3}) are not explicable by the Fan-Migdal self-energy. This points towards the possibility that strong coupling physics is at work, here.

One natural explanation, given the shape of the STS spectra in Fig.~\ref{fig:Fig3}, would be polarons in the intermediate to strong coupling regime. Within the mean-field appro\-xi\-mation, the polaron is a self-consistent localization of additional charges in the system, and this localization can be captured by DFT when self-interaction errors are eliminated~\cite{Sio2019}. Our downfolded-model approach incorporates DFT and DFPT, closely relates to the approach of Ref.~\citenum{Sio2019} in the case of a single electron being added to the system and by construction contains no self-interaction terms.

For a single electron being doped into a SL \mo{} supercell up to sizes of $18\times 18$, we did not find a localized polaron state---in line with Ref.~\citenum{Sio2023}. The situation changes, however, with more than one electron being doped into the supercell: At four or more additional electrons in an $18\times 18$ supercell, we find that n-doped SL \mo{} is unstable towards multipolaronic deformations, as can be seen in the relaxed structure depicted in Fig.~\ref{fig:Fig4}~(a). The stability of the deformation depends on the amount of doping in the system, see Appendix~\ref{metastable}.
In Fig.~\ref{fig:Fig4}~(b) we show the total energy as a function of displacement for the multipolaronic deformation of four additional electrons in the supercell, revealing an energy gain of up to \SI{25}{\milli\eV} per additional electron in the supercell. This energy gain corresponds to the formation energy when doping charges localize~\cite{Lafuente-Bartolome2022_2}, giving a total binding energy of about \SI{100}{\milli\eV} for the multipolaronic distortion shown in Fig.~\ref{fig:Fig4}~(a).
The resulting displacement pattern as well as the corresponding multipolaronic binding energy are shown to be essentially independent of the supercell size as long as the displacement pattern can be accommodated in the supercell, see Appendix~\ref{supercell} for calculations with four additional electrons on supercell sizes $6 \times 6$, $12 \times 12$ and $18 \times 18$.

To understand which phonons contribute to the multipolaronic deformation found in our lattice relaxations, we calculated the structure factor $\langle S(\mathbf q) \rangle$ (see Fig.~\ref{fig:Fig4}~(c))~\cite{Schobert2024}. It represents the Fourier transform of the atomic structure, provides an estimation of the momentum-dependent scattering intensity and encodes the polaron-associated atomic displacements. The structure factor reveals that the deformation is peaked in reciprocal space around the M-point, which corresponds to the ($2\times2$) spot with res\-pect to the reciprocal \mo~lattice. As to be expected for a localized distortion pattern, there is a finite range of $\mathbf k$-points contributing with different weights, with the ratio of the contribution from K compared to M corresponding to about 30\%. A decomposition of the displacement pattern into normal modes (see Appendix~\ref{decomposition}) shows that the LA mode dominates the polaronic displacements at the distortion-relevant $\mathbf k$-points. The phonon dispersion shown in Fig.~\ref{fig:Fig4}~(d) reveals a flat phonon band at the Brillouin zone edge, with the LA mode flattening between K and M at around \SI{28}{\milli\eV}. This phonon ener\-gy agrees well with the experimental observation of the bosonic energy of $\Omega_0 = \SI{24 \pm 4}{\milli\eV}$.

To further understand the implications of the multipolaronic deformations, we examined the corresponding electronic structure. In Fig.~\ref{fig:Fig4}~(e) we depict the undistorted band structure (blue) and the unfolded band structure in the distorted phase (orange). The weights of the unfolded bands were calculated as the overlap of the electronic wavefunctions (with the same wavevector) of the deformed and the undeformed system. These overlaps are visualized on a logarithmic colorscale as well as via the size of the markers. Dashed lines indicate the corresponding Fermi level of the undistorted (blue) and the distorted system (orange). In both band structures, the dispersive states of the Q-valley are not occupied, which resembles the situation of our experiments (cf.\@ Fig.~\ref{fig:Fig1} and Ref.~\citenum{VanEfferen2022}). A close inspection of the spectral weight of the distorted structure inside the gap of the undistorted system, however, reveals localized (i.e.\@ flat in $k$-space) states with contributions from K and Q. The charge density of these states is localized precisely in the region of strong polaronic displacements as visualized in Fig.~\ref{fig:Fig4}~(a). The occurrence of in-gap localized polaronic states featuring hybridization of K- and Q-valley states is well in-line with the dominant contribution of LA M-point phonons to the polaronic relaxation: the momentum difference K$'$Q as well as KQ$'$ corresponds to $\Gamma$M (see Appendix \ref{MomentumDifferences}), and the electron-phonon coupling matrix element scattering conduction band electrons between K- and Q-points is particularly strong for the LA M-point phonon \cite{Ge2013,Roesner2014}. 

It is intriguing to note that this coupling is active and leads to a polaronic state even though the Q-point is not occupied (Fig.~\ref{fig:Fig4}~(e)). As mentioned above, the Fan-Migdal self-energy in the deformed structure does not reveal any M-point phonon satellites in such a situation (see Appendix~\ref{spectralfunction}) \cite{Garcia-Goiricelaya2019, Lafuente-Bartolome2022_2}, showing that the polaronic deformations found in the relaxations and the spectral footprint of phonons in experiment are thus a strong-coup\-ling phenomenon. While the precise energetic minimum of the dispersive band at Q will depend on variations of the doping level and the lattice constant, the suggested scenario is qualitatively robust as long as sufficiently many electrons are doped into the system (cf.\@ Appendix~\ref{metastable} for an illustration of the case with 3 additional electrons). 

\section{Discussion}

To bring our work into context, we can rule out certain polaronic scenarios based on our observations. The energetic shifts of the peaks in response to localized charges are incompatible with purely inelastic tunneling processes, which would lead to the appearance of symmetrically distributed steps or peaks in the d$I$/d$V$ signal on either side of the Fermi level (at energies $\pm n\Omega_0$)~\cite{Stipe1998,Vitali2004}. In the same vein, renormalization of the electron bands with a particular phonon mode can be excluded as well, as it would lead to mass-enhanced (flatter) bands at fixed energies relative to $E_\text{F}$, as observed for instance in surface-doped bulk \mo{} with ARPES~\cite{Kang2018}. In our case, the spectral energy shifts induced by charges within the layer, together with the statistical distribution of the peaks, strongly suggest that the two peaks closest to the Fermi energy have a different origin.

Our experimental analysis reveals that the peaks observed in STS spectra have equidistant spacings, $\Omega_0= (24\pm4)$ meV. As the spacing corresponds to the quasiparticle energy, the simplest interpretation would be that phonon modes with energy $\Omega_0$ dominate. It is also in line with our theoretical calculations, which find that the LA phonon mode, which flattens between K and M at around $28$~meV, dominates the multipolaronic displacements. A second possibility would be a contribution of (optical) phonons with an energy close to twice (or, in general, multiples of) $\Omega_0$, as suggested theoretically in Ref.~\citenum{Garcia-Goiricelaya2019} for bulk MoS$_2$. We can, however, rule out this scenario for SL \mo~as we observe peaks well above the highest energy of a single phonon mode. In addition, we find no significant contribution to the polaronic deformation stemming from optical phonon modes in our calculations.

Our calculations revealed stable multipolaronic states, i.e.\@ complexes of more than one polaron. These polaron complexes can be viewed as precursors of the CDW emerging in \mo{} under strong electron doping, where the periodic lattice distortion bases on the LA M-phonon and its strong coupling to K- and Q-electrons~\cite{Ge2013, Roesner2014, BinSubhan2021, Marini2023}. The appearance of the multipolarons in STM here is thus considerably different from polarons in 2D semiconductors such as CoCl$_2$~\cite{Liu2023,Cai2023}. There, a depression in the apparent height is observed due to the weak screening of the local charge. The interpretation of multipolarons being precursors of the charge density wave in \mo\, is consistent with the appearance of a weak local ($2\times2$) superstructure in our STM images.  We expect that further doping will lead to a global periodic lattice distortion accompanied by a ($2\times2$) charge density wave superstructure with respect to the \mo~lattice.

\subsection{Characteristics of multipolaronic spectra}

The spectral properties associated with multipolaronic states are less understood than the well-studied theoretical framework of the single polaron problem. Yet, the comparison to the better studied bipolaron case allows us to speculate about the spectral features exhibited by multipolarons. 

For the Holstein model of the simplest two-site polaron system, the spectral intensity is characterized by the emergence of a series of equidistant, phonon-induced replica peaks above and below an energy gap $E_\mathrm{gap}$ when increasing the electron-phonon-coupling strength~\cite{deMello1997}. This energy gap is equal to the binding energy $2\varepsilon_{\rm{pol}}$~between two monopolarons in the strong coupling limit, $E_\mathrm{gap}=2\varepsilon_{\rm{pol}}$, with the corresponding Fermi energy positioned precisely in the middle of the gap. The peaks below the Fermi energy correspond to electron removal from an occupied (bi)polaronic state, while the peaks above the Fermi energy are associated with electron addition to (bi)polaronic states.
These characteristics are expected to hold true generally for any interacting many-polaron system on a lattice and fit qualitatively well to the characteristics of our experimentally observed STS spectra. Similar spectra of equidistant, phonon-induced peak structures \cite{Ranninger1993} separated by an energy gap are reported in other theoretical works, such as the study of the Holstein many-polaron model via cluster perturbation theory \cite{Hohenadler2006}, the quantum Monte Carlo and exact diagonalization treatment of the 1D spinless Holstein model of a many-polaron system \cite{Hohenadler2005}, as well as the DMFT and exact diagonalization treatment of the Holstein bipolaron at half-filling \cite{Capone2006}.

Under account of the direct comparison between the bipolaron case and our proposed multipolaron case, we can thus estimate the multipolaronic binding energy as $E_\mathrm{gap}\approx40$~meV from our STS spectra, which is in the same order of magnitude as the multipolaronic binding energy of $\sim100$~meV derived from our qualitative theory comparison in Fig. \ref{fig:Fig4}~(b).

While our statistical analysis in Fig. \ref{fig:Fig2}~(a) shows a preference for symmetric peaks around the Fermi energy, in line with the expected spectral features of multipolaronic states, clear energy shifts are observed in the vicinity of local charges. For the spectra in the vicinity of Cs adsorbates (see Fig. \ref{fig:Fig3} and \ref{fig:linescan}), the entire series of spectral peaks shifts rigidly in the vicinity of Cs without strong changes in the gap size. Due to charge transfer from Cs to \mo~and the underlying graphene, we expect strong electric fields around the Cs atoms. We suspect that band bending due to these electric fields is the reason behind the shifts in the spectra seen in the Cs-decorated case, while the energy gap remains essentially unchanged. Based on these observations, the large scatter in $E_{\rm{gap}}$ as seen in Fig. \ref{fig:Fig2}~(c) is likely not caused by charge inhomogeneities. Possible origins are structural disorder~or variations in the intercalation layer.

\subsection{Limits and plausibility of the downfolding approach}
It is important to justify the use of the downfolding approach in our theoretical analysis. The proposed multipolaronic states might still be altered by electron-electron interaction effects which are neither included in the Holstein model of (bi)polarons nor in our downfolding model: non-local exchange enhances electron-phonon coupling vertices and could stabilize polaron complexes involving less than the proposed four electrons. At the same time, electron-electron interactions can lead to correlation effects beyond the effective mean field treatment of the polaronic state considered here. 

In free-standing 2D semiconductors as well as 2D semiconductors in a weakly screened environment, long-range Coulomb interactions could counteract the proposed multipolaron state, since the formation of bipolarons and by extension multipolarons is limited to a certain range of the ratio between the electron-phonon coupling and the Coulomb repulsion interaction strength \cite{Verbist1992, Frank2010}. Therefore, for strong Coulomb interaction, the formation of multipolarons is suppressed.

Though strong Coulomb interactions are observed in semiconducting \mo{} when charges localize within grain boundaries~\cite{Jolie2019,vanEfferen2024}, here we are adding charge carriers into the conduction band of n-doped \mo~on top of a metallic substrate. In consequence, the \mo/substrate system is conducting and can effectively screen charges within the layer. This is experimentally confirmed as a reduced extension of band bending effects in real space as compared to semiconducting \mo{} \cite{VanEfferen2022}. In consequence, long-range Coulomb interactions emerging from the formation of multipolaronic states will be effectively screened in our metallic system. So it stands to reason that the possibility of multipolaron formation should be further supported.

While a full account of Coulomb effects remains to be given, we also note that for the downfolding approach used here, many contributions of interaction terms are shown to be canceling out \cite{Schobert2024}. Due to that, the Born-Oppenheimer potential energy surfaces of a variety of metallic charge density wave materials have been reproduced to a high degree of accuracy. Effectively, the \mo~sample equals a “metallic” setup with quenched long-range interactions here, which could explain the qualitative agreement regarding the binding energies between our calculations and the experiment.

These arguments provide evidence that our proposed scenario of multipolaronic binding concluded from our downfolding calculations is a reasonable assumption to make, given the clear spectral similarities to literature without the full account of electron-electron interaction and regarding the validity of our downfolding approach in metallic charge density wave materials.

To the best of our knowledge, the problem of electronic spectra resulting from multipolaronic states under full account of the electron-electron interaction has not been addressed for real materials so far, and we hope that our study triggers a closer theoretical investigation.

\section{Summary}
In summary, our results demonstrate that multipolaronic states emerging from the coupling of charge carriers to a well-defined phonon mode can be measured via STM in n-doped metallic SL \mo. A local superstructure resembling a ($2\times2$) is observed experimentally. In d$I$/d$V$ spectra, two strong resonances are observed on either side of the Fermi level, together with inelastic phonon replica evenly spaced by an energy of $\Omega_0 = \SI{24 \pm 4}{\meV}$. By combining density functional (perturbation) theory with a recently developed \textit{ab initio} electron-lattice downfolding technique, we find that the \mo{} lattice has a strong tendency towards localized deformations with a finite range of wave vectors around the M point contributing to the distortion, which can be interpreted as a multipolaron involving four (or more) electrons. By revealing the effects of large electron-phonon coupling on the tunneling spectra of 2D materials, we expect that our findings create new research opportunities into 2D (multi-)polaron complexes.

\section{Acknowledgements}
We acknowledge funding from the Deutsche Forschungsgemeinschaft (DFG, German Research Foundation) through CRC 1238 (project No. 277146847, subprojects A01 and B06). J.F. acknowledges financial support from the DFG through project FI 2624/1-1 (project No. 462692705) within the SPP 2137. J. B. acknowledges financial support from the DFG through Germany's Excellence Strategy (University Allowance, EXC 2077, project No. 390741603, University of Bremen). Financial support by the DFG priority program SPP2244 is acknowledged by M.W., T.W. (project No. 43274199) and W.J. (project No. 535290457). L.P. and T.W. thank the DFG for funding through the research unit QUAST FOR 5249 (project No. 449872909; project P5). A.S., Y. in 't V. and T.W. acknowledge funding and support from the DFG through the cluster of excellence ``CUI: Advanced Imaging of Matter'' of the Deutsche Forschungsgemeinschaft (DFG EXC 2056, project No. 390715994). The authors gratefully acknowledge the computing time made available to them on the high-performance computer "Lise" at the NHR Center NHR@ZIB under the project hhp00063. This center is jointly supported by the Federal Ministry of Education and Research and the state governments participating in the NHR (\url{www.nhr-verein.de/unsere-partner}).

\appendix
\section{Experimental methods}
\label{methods}

The samples were grown \textit{in situ} in two preparation chambers with base pressure $p < 5 \times 10^{-10}$ mbar. Ir(111) is cleaned by \SI{1.5}{\kilo\eV} Ar$^+$ ion sputtering and annealing to a temperature of \SI{1550}{\K}. Gr is grown on Ir(111) in two steps. First, the Ir is exposed to ethylene until saturation, followed by \SI{1370}{\K} thermal decomposition in order to obtain well-oriented Gr islands. Afterwards, the sample is exposed to \SI{2000}{\liter} ethylene at \SI{1370}{\K} for \SI{600}{\s}, which yields a complete single-crystal Gr layer~\cite{Coraux2009}. SL \mo{} is grown by Mo deposition in an elemental S pressure of $\SI{1 \times 10^{-8}}{\milli\bar}$~\cite{Hall2018}. Subsequently, the sample is annealed to \SI{1050}{\K} in the same S background pressure. Cs was evaporated using a direct current evaporator to heat a Cs dispenser in ultra-high vacuum. During evaporation, the \mo/Gr/Ir(111) sample was kept at \SI{570}{\K}. After intercalation, the sample was flashed to \SI{730}{\K} to remove residual Cs from the surface. Details about the Eu intercalation can be found in Ref.~\citenum{VanEfferen2022}.

STM and STS were carried out in two ultra-high vacuum systems at a temperature of \SI{0.4-6.5}{\K} after \textit{in situ} transfer of the samples from the preparation chamber.

\begin{figure}[tb]
	\centering
		\includegraphics[width=0.45\textwidth]{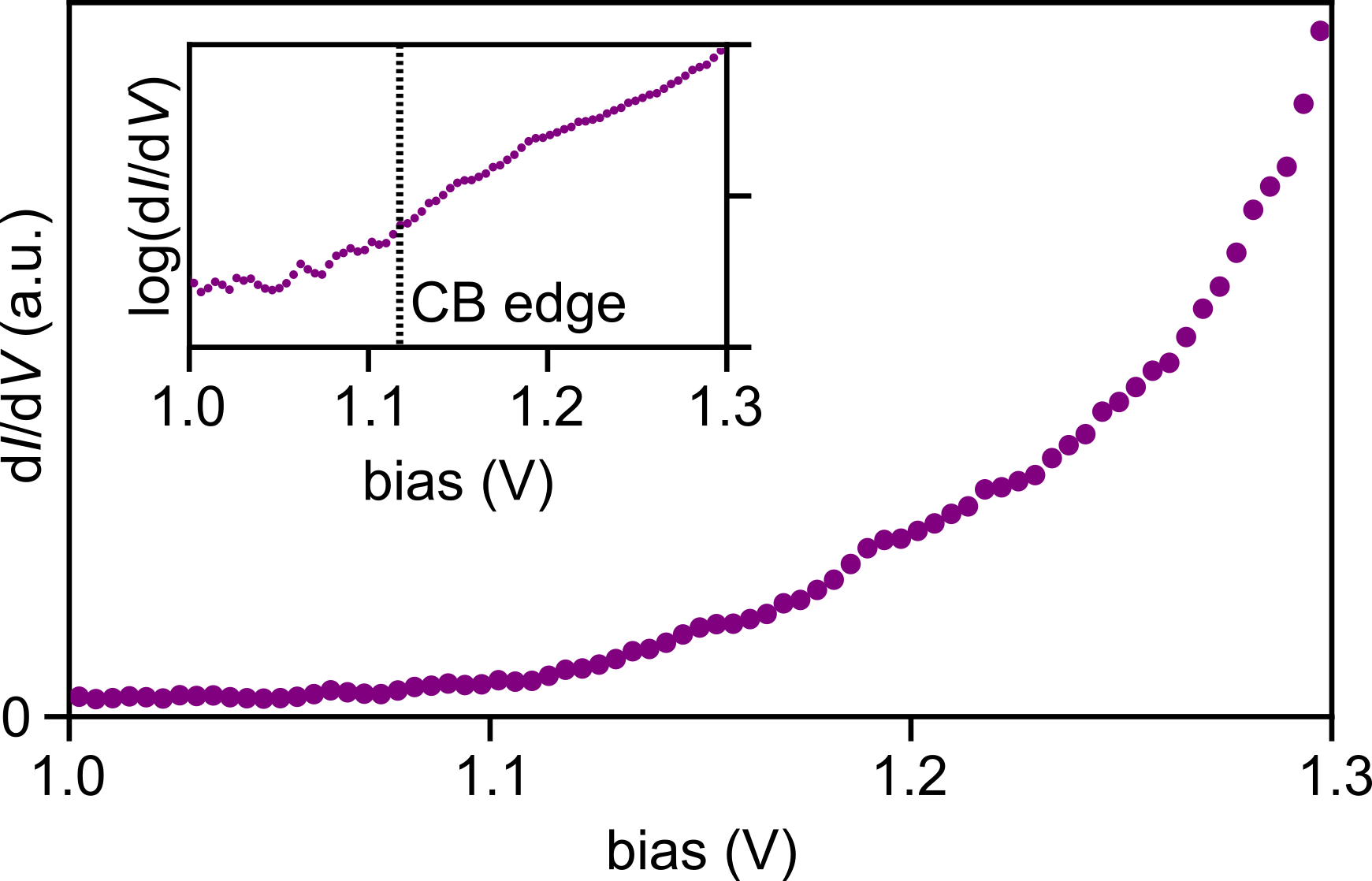}
			\caption{Absence of peak-dip features in semiconducting \mo.~In semiconducting \mo, here on a Gr/O/Ir(111) substrate, the band edge is smooth. No peak-dip features are seen.
									STM/STS parameters: $V_\text{set} = \SI{1.35}{\V}$, $I_\text{set} = \SI{1.0}{\nA}$, $V_\text{mod} = \SI{5}{\mV}$.}
 \label{fig:FigSCB}
\end{figure}

\section{Calculations details}
\label{sec:calculation_details}

DFT calculations were performed with \textsc{Quantum ESPRESSO}~\cite{Giannozzi2009} using the PBE approximation for the exchange-correlation potential~\cite{Perdew1996} and norm-conserving pseudo-potentials from the PseudoDojo database \cite{Hamann2013, vanSetten2018}. Plane waves until an energy cutoff of \SI{100}{Ry} were included and a Fermi-Dirac-type smearing of \SI{0.001}{Ry} was imposed. We considered a doping of $\SI{0.01}{e^{-}}$/MoS$_2$, which corresponds to a dopant density of $1.14\times 10^{13}$\,\si{\per\centi\meter\squared}. We included a vacuum of approximately \SI{12}{\smallskip}\AA~above the monolayer. We used an $18 \times 18$ $\mathbf k$-point grid for the unit cell. Relaxation of the (un)doped unit cell for fixed cell height yielded the lattice constant of \SI{3.23}{\smallskip}\AA{} (\SI{3.19}{\smallskip}\AA). We used these unit cell structures and their lattice constants as the starting point of all subsequent calculations.

For the DFPT calculations via \textsc{Quantum ESPRESSO}, we used a $6 \times 6$ $\mathbf q$-point grid. For the determination of the electron-phonon coupling we employed the \textsc{EPW} code \cite{Ponce2016}. Three energy bands around the Fermi energy were transformed into the Wannier basis starting from orbitals of Mo $d_{z^2}$-, $d_{x^2-y^2}$-, and $d_{xy}$-character.
Spectral function calculations in Appendix~\ref{spectralfunction} were carried out on top of the three band Wannier model extended to include spin-orbit coupling within the EPW code including possible Fan-Migdal contributions to the self-energy. A dense $\mathbf q$-point grid (192$\times$192) was used. Smearing in electron-phonon coupling summation and energy-conserving delta functions was set to \SI{0.002}{\eV}.

The relaxations on large supercells ($18 \times 18$) were facilitated by the downfolding strategies from Ref.~\citenum{Schobert2024}, which have been implemented in the \textsc{elphmod} package~\cite{Berges2017}. The same parameters as in DFT were used except for the $\mathbf k$-point grid, which corresponds to a $\Gamma$-only calculation on the $18 \times 18$ supercell.

We used the ``model III'' variant of the downfolding strategy of Ref.~\citenum{Schobert2024}, where its equivalence to DFT had been established. Full equivalence would require recalculation of model parameters (charge self-consistency). The Landau-Pekar model provides a formally equivalent description of DFT polarons, with the equivalence becoming exact in the Fr\"ohlich limit~\cite{Sio2019}. In the Fr\"ohlich model, the polaronic self-energy within the self-consistent many-body approach provides the Landau-Pekar solution to the Fr\"ohlich polaron, and thus at least the formal solution to the polaron problem in DFT and in downfolding model III~\cite{Lafuente-Bartolome2022_2}. Thus our approach captures localization and static renormalization effects encoded in the polaronic self-energy. The Fan-Migdal self-energy adds dynamical renormalization of the energy levels.

\section{Absence of peak-dip features in semiconducting \mo}
\label{absence}

We only observed the peak-dip features in d$I$/d$V$ spectra measured on metallic \mo. Fig.~\ref{fig:FigSCB}~shows a d$I$/d$V$ spectrum measured on \mo/Gr/O/Ir(111), close to the conduction band minimum. Oxygen is known to shift the conduction band of \mo~to higher energies, making \mo~semiconducting~\cite{VanEfferen2022}. The band edge is smooth, without pronounced additional features. The absence of peak-dip features is in line with the absence of polarons in semiconducting \mo.

\begin{figure}[t]
	\centering
		\includegraphics[width=0.45\textwidth]{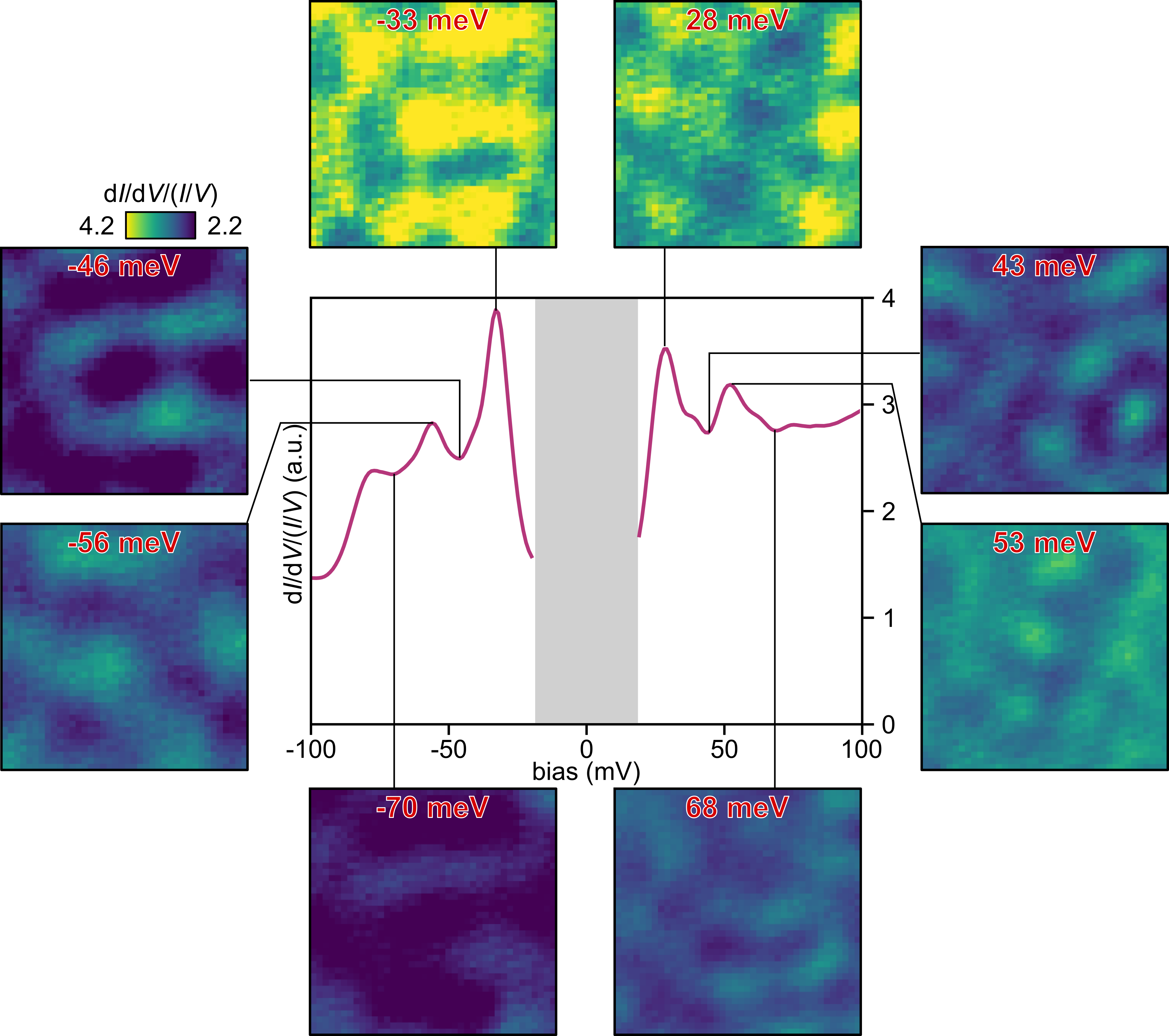}
			\caption{Delocalized quasiparticle scattering in metallic \mo.~Conductance maps of a large, defect free area of \mo. The normalized spectrum in the center is spatially averaged over the whole investigated area. While the characteristic resonances are visible, they are delocalized, leading to an increase or decrease of the conductance at any point in the area. The patterns that are visible are related to quasiparticle interference in \mo.
						STM/STS parameters: \SI[parse-numbers=false]{10 \times 10}{\nano\meter\squared}, $V_\text{set} = \SI{0.1}{\V}$, $I_\text{set} = \SI{1.0}{\nA}$, $V_\text{mod} = \SI{2}{\mV}$, $f_\text{mod} = \SI{877}{\hertz}$.
			}
 \label{fig:FigSRealSpace}
\end{figure}

\begin{figure}[ht]
	\centering
		\includegraphics[width=0.4\textwidth]{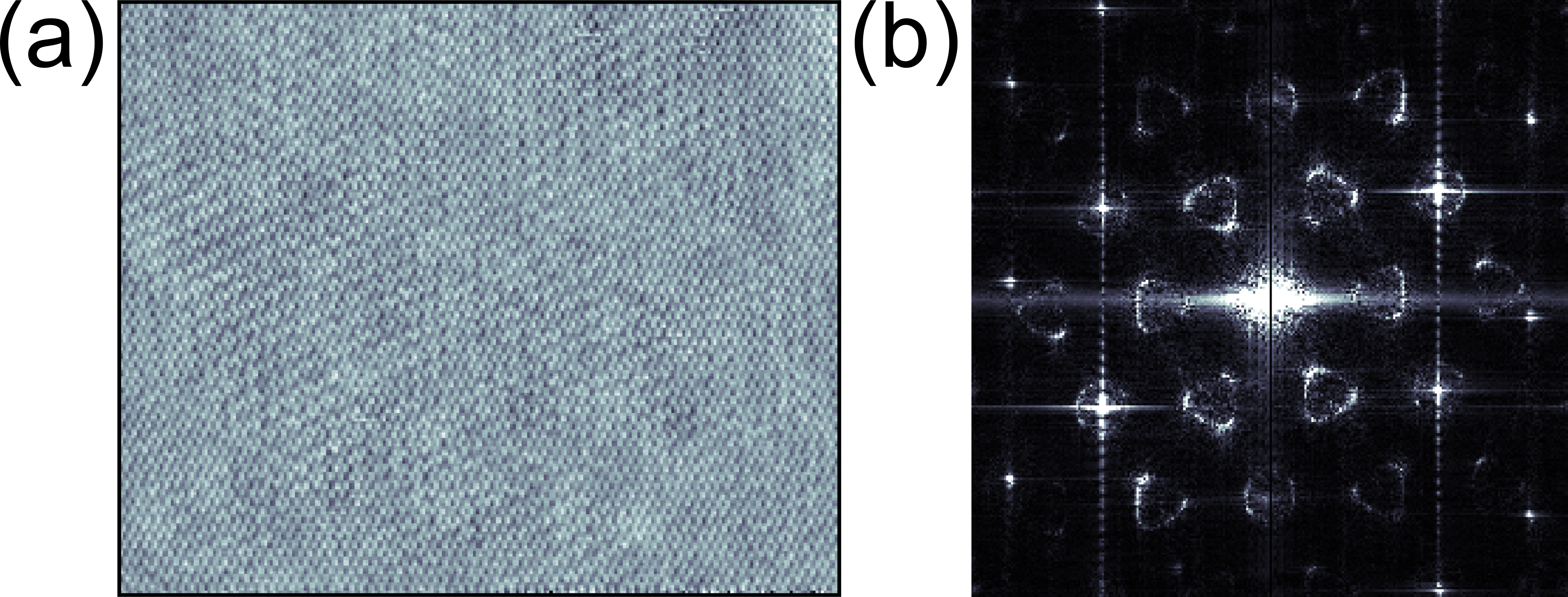}
			\caption{Scattering pattern in Cs-intercalated graphene. (a)~Conductance (d$I$/d$V$) map and (b)~corresponding Fourier transformation of Cs-doped graphene.
									STM/STS parameters: (a)~\SI[parse-numbers=false]{17 \times 14}{\nano\meter\squared},~$V_\text{set} = \SI{25}{\mV}$, $I_\text{set} = \SI{1.0}{\nA}$, $V_\text{mod} = \SI{25}{\mV}$.}
 \label{fig:FigSCs}
\end{figure}

\begin{figure*}[tb]
	\centering
		\includegraphics[width=0.85\textwidth]{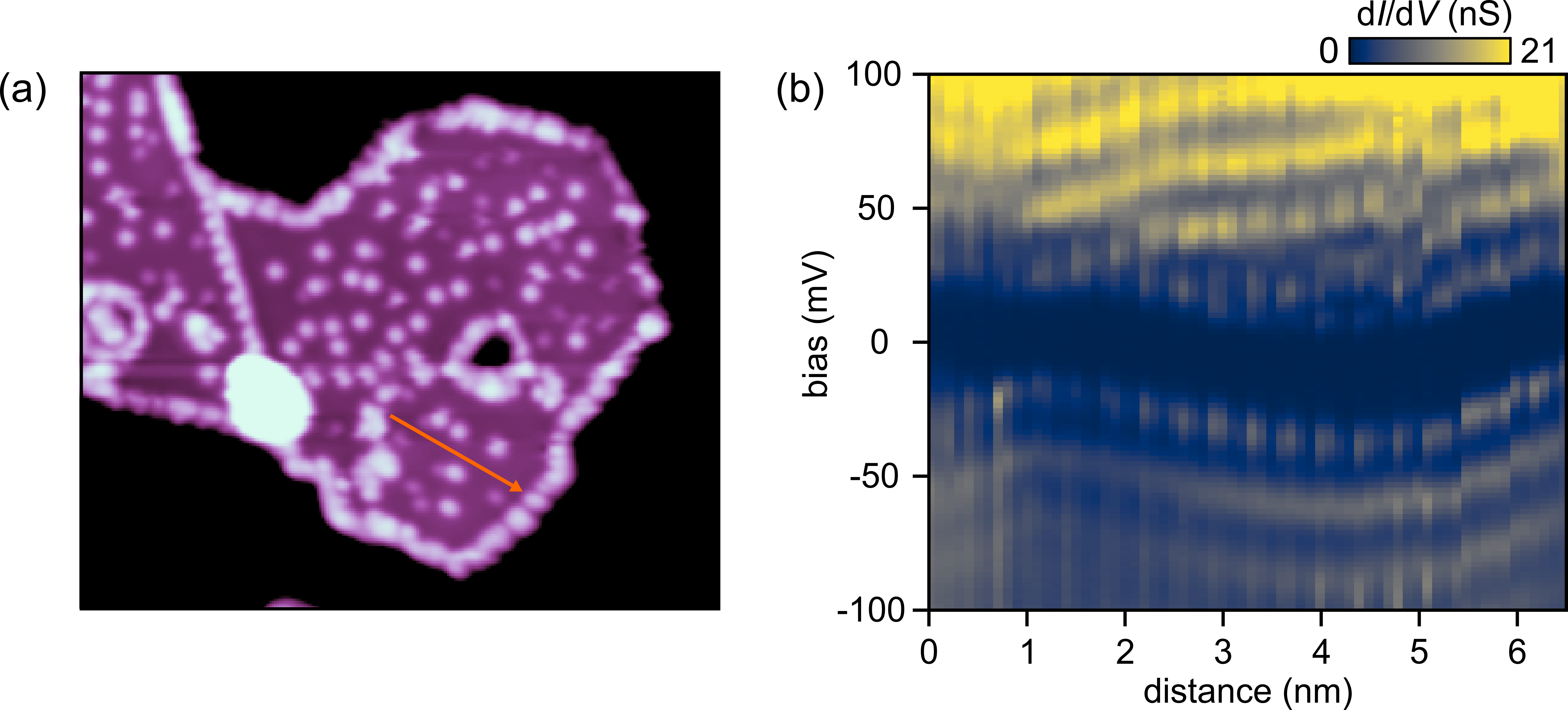}
			\caption{Effect of an inhomogeneous electrostatic environment. (a)~STM image of a SL \mo{} island on Gr/Cs/Ir(111) with adsorbed Cs atoms. The arrow marks the position of data acquisition. (b)~d$I$/d$V$ intensity as a function of position and bias voltage.
			STM/STS parameters: (a)~\SI[parse-numbers=false]{40 \times 33}{\nano\meter\squared}, $V_\text{set} = \SI{1.0}{\V}$, $I_\text{set} = \SI{100}{\pA}$; (b)~$V_\text{set} = \SI{100}{\mV}$, $I_\text{set} = \SI{1.1}{\nA}$, $V_\text{mod} = \SI{2}{\mV}$, $f_\text{mod} = \SI{877}{\Hz}$.
			}
 \label{fig:linescan}
\end{figure*}

\section{Delocalized quasiparticle scattering in metallic \mo}
\label{QPI}

Standing waves are observed when mapping the d$I$/d$V$ signal in real space, see Fig.~\ref{fig:FigSRealSpace}. These are related to quasiparticle interference (QPI), arising from electron scattering near the \mo~K point~\cite{VanEfferen2022}. A d$I$/d$V$ spectrum spatially averaged over the entire area under investigation still exhibits pronounced peak-dip features, see Fig.~\ref{fig:FigSRealSpace}. This ensure that these peak-dip features are not related to local density variations due to QPI.

\section{Doping level of Cs intercalated sample}
\label{doping}

In the conductance map of graphene in Fig.~\ref{fig:FigSCs}~(a), standing wave patterns can be observed, which arise from QPI of the graphene electrons. A Fourier transform, shown in Fig.~\ref{fig:FigSCs}~(b), reveals triangular scattering patterns around the K- and K$'$-points. The size of these triangular pockets is determined by intervalley scattering within the Dirac cones and is thus directly linked to the constant energy contours of the graphene bandstructure at the energy of the measurement~\cite{Dombrowski2017}. At \SI{25}{\meV}, we find that the width of the trigonally warped Dirac cone, measured from $\Gamma \rightarrow \rm{K} \rightarrow \rm{M}$ is \SI{0.374}{\angstrom^{-1}}. Comparing this value to the fully Cs-intercalated data of Ref.~\citenum{Dombrowski2017}, we estimate that $E_\text{D} \approx \SI{-0.9}{\eV}$, with $E_\text{D}$ the Dirac point of graphene. Compared to the Dirac point of graphene in the undoped \mo/Gr/Ir(111) system ($E_\text{D} = \SI{0.25}{\eV}$)~\cite{Ehlen2018}, this shift is in considerable excess of the energy needed to bring the \mo{} conduction band below the Fermi level~\cite{Murray2019, VanEfferen2022}.

\begin{figure}[htb]
	\centering
		\includegraphics[width=0.41\textwidth]{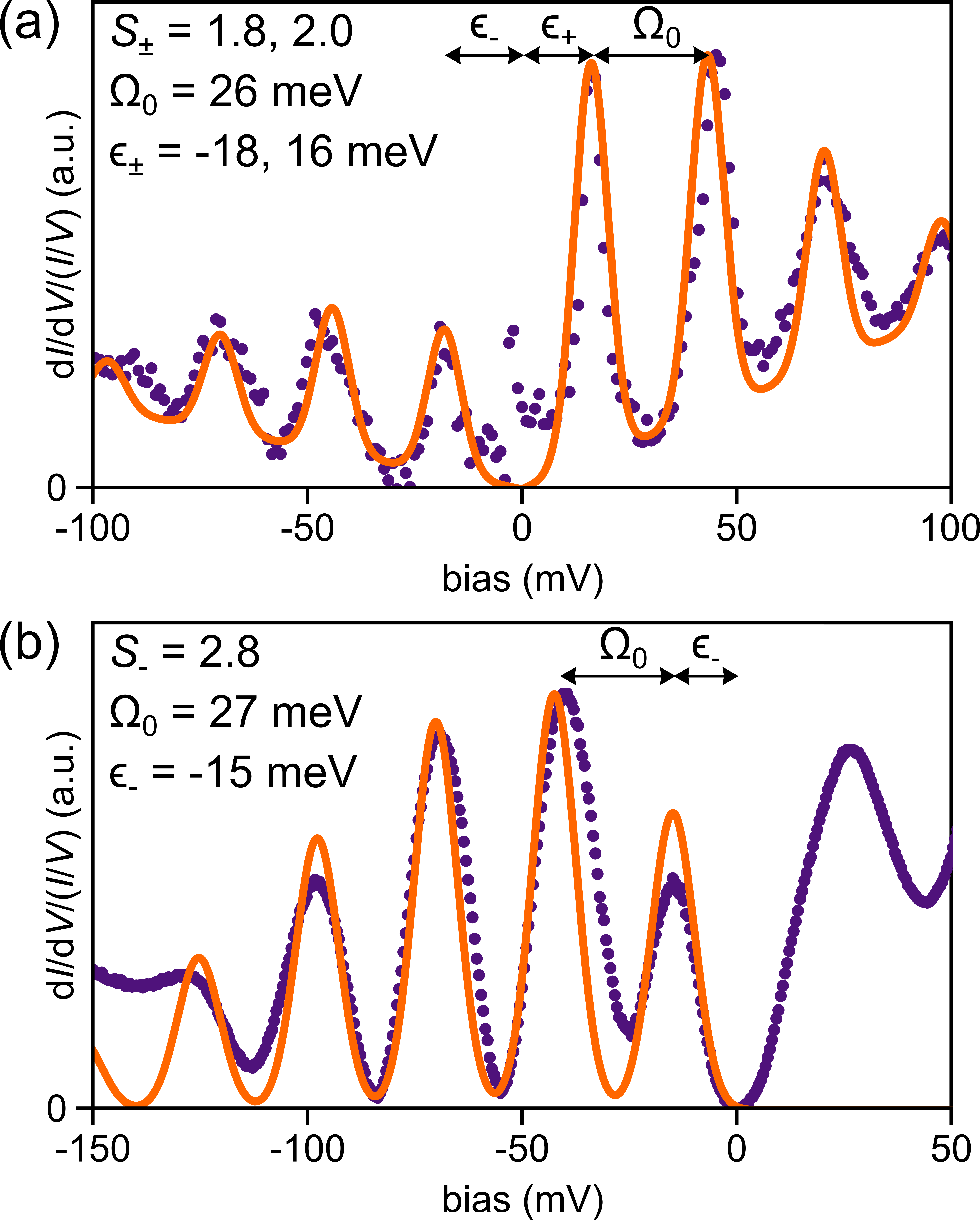}
			\caption{Variation in the inelastic tunneling processes. (a)-(b)~Two conductance spectra taken on defect-free areas of \mo/Gr/Eu/Ir(111) are shown. The spectra are fitted using the model of Eq.~1 in the main manuscript~\cite{Cochrane2021}. 
									STM/STS parameters: (a)~$V_\text{set} = \SI{0.1}{\V}$, $I_\text{set} = \SI{500}{\pA}$, $V_\text{mod} = \SI{1}{\mV}$; (b)~$V_\text{set} = \SI{0.1}{\V}$, $I_\text{set} = \SI{500}{\pA}$, $V_\text{mod} = \SI{5}{\mV}$.}
 \label{fig:FigSFitting}
\end{figure}

\begin{figure}[tb]
	\centering
		\includegraphics[width=0.45\textwidth]{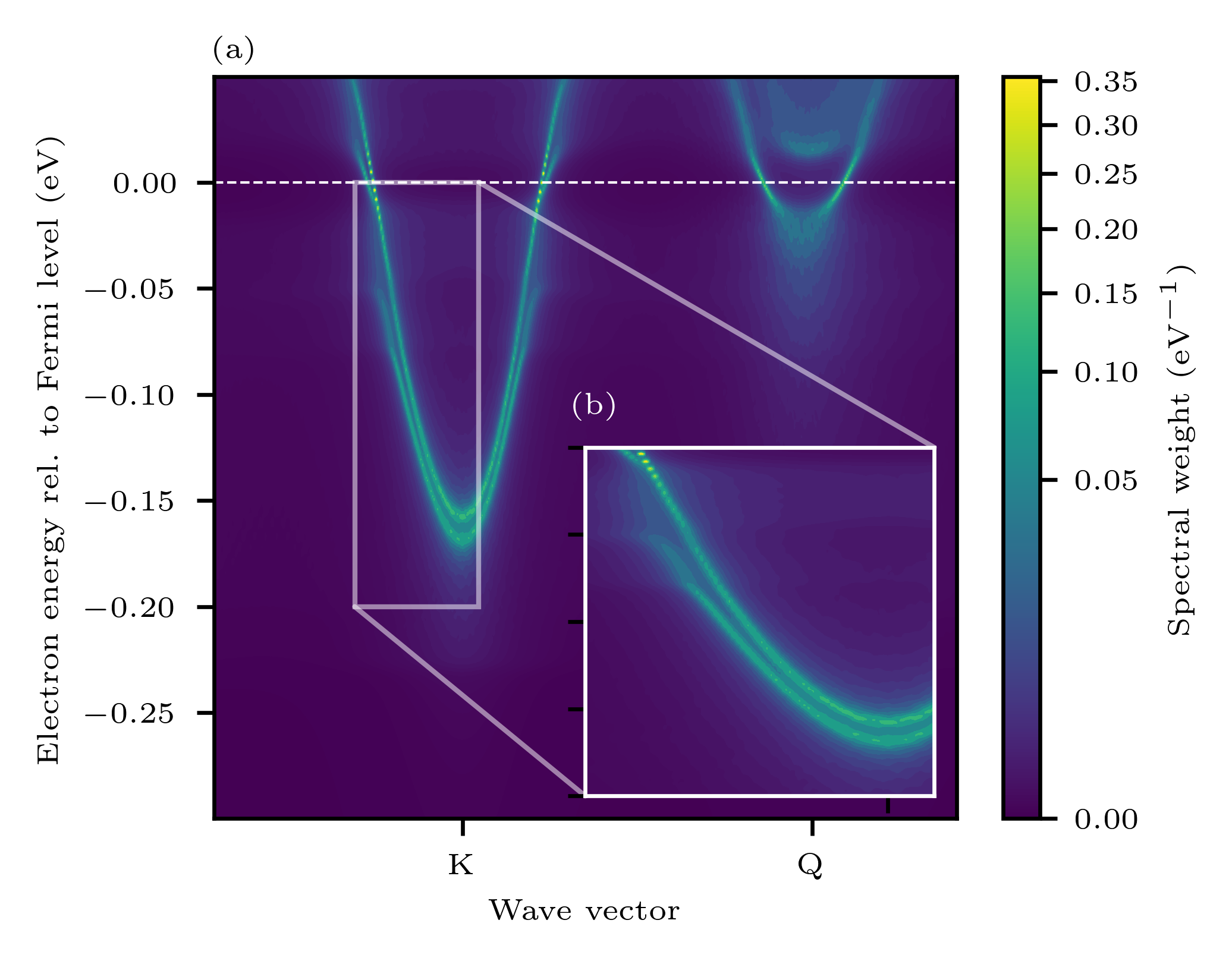}
        \includegraphics[width=0.45\textwidth]{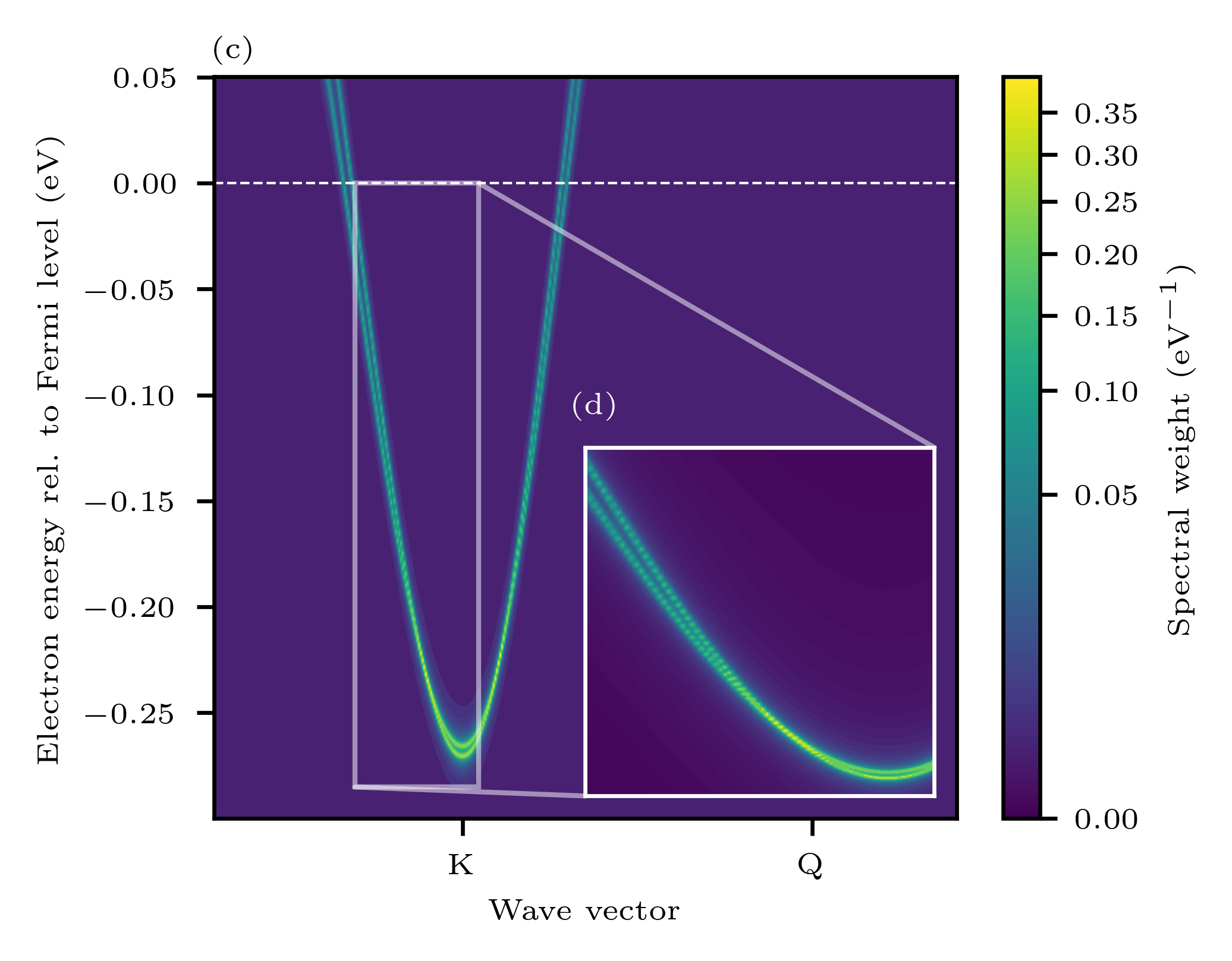}
			\caption{Conduction band spectral function with Fan-Migdal self-energy at different valley occupations. (a),\,(b) K- and Q-valleys are occupied; (c),\,(d) Only K-valleys but not Q valleys are occupied. Hence, the phonon-induced kinks and band cuts visible in (b) are no longer realized in (d).
			}
    \label{fig:FigSSpectralFunctions}
\end{figure}

\section{Spectra measured close to Cs atoms on \mo}
\label{Cslinescan}
Additional spectra measured close to Cs atoms on \mo~are presented in Fig.~\ref{fig:linescan}.  They are recorded along the line in Fig.~\ref{fig:linescan}~(a) and shown as color plot in Fig.~\ref{fig:linescan}~(b), which visualizes the d$I$/d$V$ intensity as a function of position and bias voltage.

Again we find that the peaks shift considerably in energy, while not crossing the Fermi energy. Compared to Fig.~\ref{fig:Fig3}~(b), the behavior of the spectra is shown to be more complex. This behavior is likely due to the highly complex combination of (multi-)polaron physics and the inhomogeneous electrostatic environment close to the Cs adatoms.

\section{Electron-phonon coupling strength from conductance spectra}
\label{Sfitting}

Fig.~\ref{fig:FigSFitting}~shows additional spectra measured on \mo. We observe a variation in the intensity of the peak-dip features. This variation leads to a scatter in the Huang-Rhys factor $S_\pm$. The phonon energy $\Omega_0$ is not affected by the intensity variations. A constant phonon energy is in line with our interpretation that one phonon mode dominates the spectral intensity.

\section{Spectral function with Fan-Migdal self-energy contributions and cumulant expansion}
\label{spectralfunction}

From the analytical expression for the Fan-Migdal self-energy and for the spectral function (see, e.g., Ref.~\citenum{Garcia-Goiricelaya2019}) we can understand the presence/absence of phonon-induced kinks and band cuts in the electronic K-valley depending on whether K- and Q-valleys or only the K-valleys are occupied by dopant electrons (see Fig.~\ref{fig:FigSSpectralFunctions}):

\begin{widetext}
\begin{equation}
    \Sigma_{\mathbf k j}(\omega) = \frac{1}{N_\mathbf q} \sum_{\mathbf q i \nu} \left| g_{ij}^\nu (\mathbf k, \mathbf q) \right|^2 \left( \frac{f\left(\varepsilon_i^{\mathbf k + \mathbf q}\right)}{\omega - \left(\varepsilon_i^{\mathbf k + \mathbf q} - \omega_\nu^{\mathbf q}\right) + \mathrm i \eta} + \frac{1 - f\left(\varepsilon_i^{\mathbf k + \mathbf q}\right)}{\omega - \left( \varepsilon_i^{\mathbf k + \mathbf q} + \omega_\nu^{\mathbf q}\right) + \mathrm i \eta} \right)
\end{equation}

\begin{equation}
    A \left( \mathbf k, \omega \right) = -\frac 1\pi \sum_j \Im \left( G \left( \mathbf k, \omega \right) \right) = -\frac 1 \pi \sum_j \frac{ \Im\left( \Sigma_{\mathbf k j}(\omega) \right) }{\left( \omega - \varepsilon_j^{\mathbf k} - \Re \left( \Sigma_{\mathbf k j}(\omega) \right) \right)^2 + \left( \Im \left( \Sigma_{\mathbf k j}(\omega) \right) \right)^2}
\end{equation}
\end{widetext}

\begin{figure*}[tb]
	\centering
		\includegraphics[width=1\textwidth]{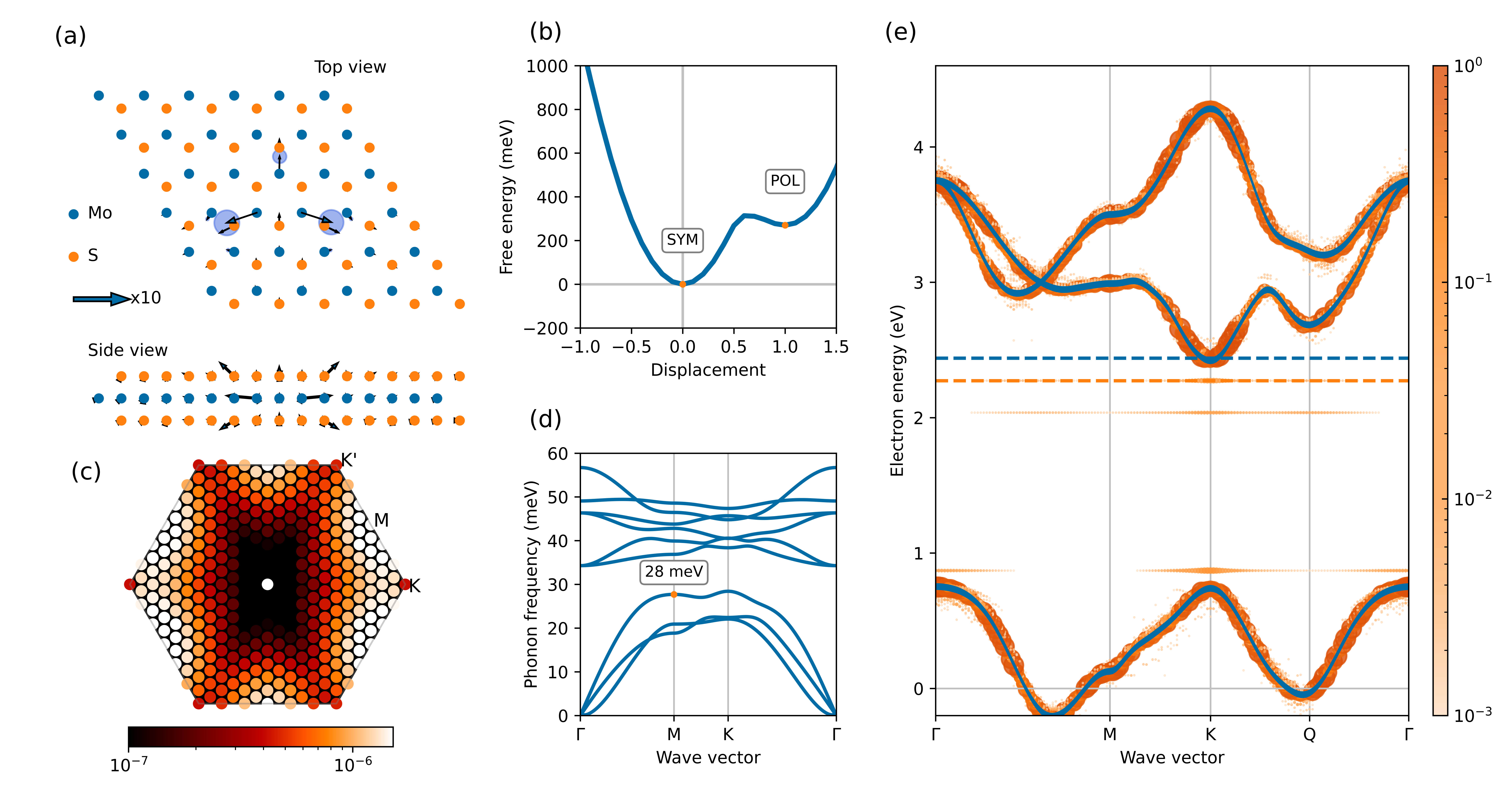}
			\caption{Doping dependence of polaronic deformations.~This figure was created in analogy to Fig.~4 of the main text, see there for more detailed information on the subfigures. ~(a) Zoom into relaxed crystal structure based on \textit{ab initio} model calculations on an $18 \times 18$ supercell with a lattice constant of $\SI{3.19}{\AA}$ at an electronic doping of $\SI{0.0093}{e^{-}}$ per formula unit of \mo{}. The doping results in 3 additional electrons per supercell and blue circles indicate the localization of these additional electrons. We find a metastable polaronic deformation and a lowering of the symmetry of the system. (b)~Total energy as function of displacement for the metastable polaronic deformation shown in (a). (c)~Structure factor $S(\mathbf q)$ for the relaxed geometry seen in (a), which also shows the lowering of the symmetry. (d)~Phonon bands obtained via DFPT at an electronic doping of $\SI{0.01}{e^{-}}$/MoS$_2$. (e)~Band structure of low-energy subspace as obtained by DFT (blue) and after the \textit{ab initio} model relaxation (orange). The polaron formation is accompanied by the appearance of dispersionless states deep inside the gap which pull down the Fermi level (dashed lines). 
			}
   
 \label{fig:MetastablePolaron}
\end{figure*}

The self-energy $\Sigma$ has poles at $\omega = \varepsilon_i^{\mathbf k + \mathbf q} - \omega_\nu^\mathbf{q} - \mathrm i \eta$ in the first fraction, which affect the occupied part of the spectrum due to the Fermi function in the numerator $f\left(\varepsilon_i^{\mathbf k + \mathbf q}\right)$. Due to the factor $1-f\left(\varepsilon_i^{\mathbf k + \mathbf q}\right)$, the second term in $\Sigma$ only acts on the unoccupied part of the spectrum. Therefore, the band cuts and kinks below the Fermi level originating from phonon-induced scattering between K- and Q-valleys can only be realized if the Q-valley is occupied. A visual representation is given in Fig.~\ref{fig:FigSSpectralFunctions}.

The Fan-Migdal self-energy fails to reproduce the full series of polaronic replica in the spectral function even in systems with strong polarons. Using the cumulant expansion approach (similar to Refs.~\citenum{Ulstrup2024,Kas2014}), a full cascade of replica can be obtained. In the framework of the cumulant expansion, the Green's function can be written as 
\begin{equation}
    G(\mathbf k,t) = - \mathrm i Z_{\mathbf k} \Theta (t>0) e^{\mathrm i (-\varepsilon_{\mathbf k}+\mu
    + \Delta_{\mathbf k} + \mathrm i \eta)t}e^{O_{\mathbf k} (t)}\,,
\end{equation}
where
\begin{equation}
    O_{\mathbf k}(t) = \int\limits_{-\infty}^{\infty} \mathrm d \omega \frac{\vert \Im (\Sigma (\mathbf k,\omega + \varepsilon_{\mathbf k}))\vert }{\pi \omega^2} e^{-\mathrm i \omega t}
\end{equation}
captures the induction of shakeoff bands,
\begin{equation}
    \Delta_{\mathbf k} = \int\limits_{-\infty}^{\infty} \mathrm d \omega \frac{\vert \Im (\Sigma (\mathbf k,\omega + \varepsilon_{\mathbf k}))\vert }{\pi \omega}
\end{equation}
shifts the quasiparticle dispersion, and
\begin{equation}
    Z_{\mathbf k} = \exp\left( - \int\limits_{-\infty}^{\infty} \mathrm d \omega \frac{\vert \Im (\Sigma (\mathbf k,\omega + \varepsilon_{\mathbf k}))\vert }{\pi \omega^2} \right)
\end{equation}
acts as a renormalizing factor. By construction, to first order in the perturbation strength the Fan-Migdal self-energy and the cumulant expansion must give the same contribution. $Z_{\mathbf k}$ can thus be extracted from the \textit{ab-initio} Fan-Migdal self-energy. In the case of occupied K-valley and unoccupied Q-valley, we obtain $Z_K \approx 0.9$\,. Therefore, the maximal redistributable spectral weight in a cumulant expansion treatment is bound to approximately 10\,\%\,. This goes against the experimental observations depicted in Fig.~\ref{fig:Fig3}\,(c), where spectral weight redistribution is much larger than 10\,\%\,. In conclusion, neither the Fan-Migdal self-energy nor a cumulant expansion treatment does explain the polaronic replica seen in experiment. Thus, a monopolaron explanation can be ruled out here.

\begin{figure*}[tb]
	\centering
		\includegraphics[width=1\textwidth]{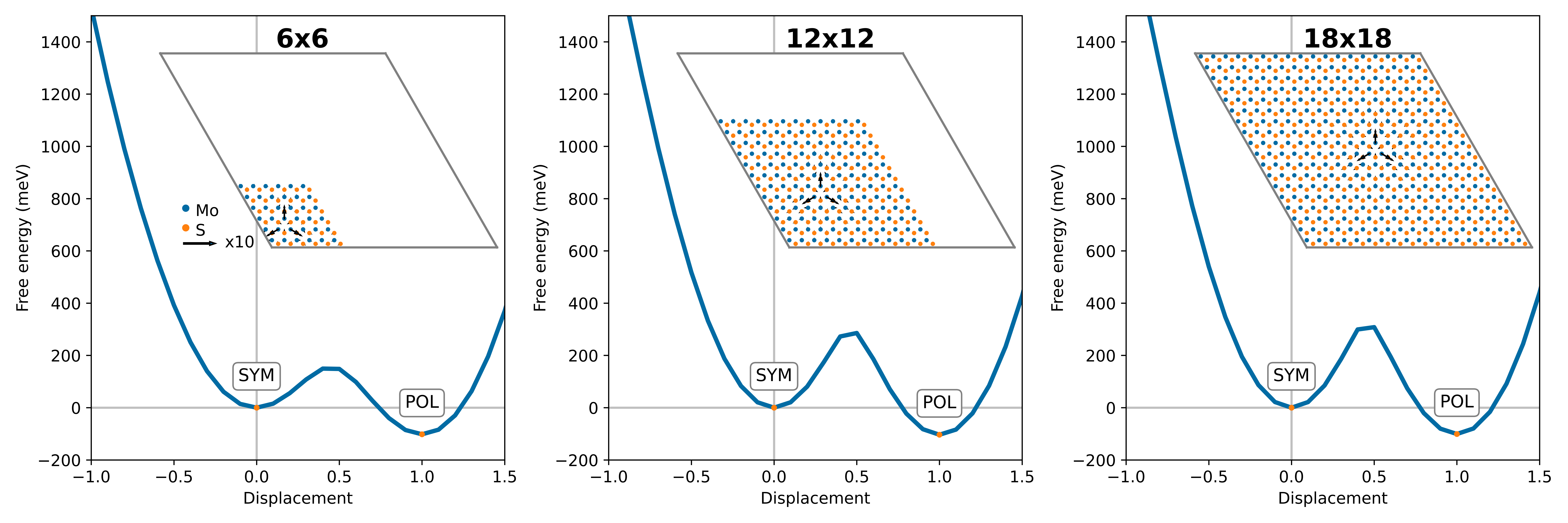}
			\caption{Dependence of the displacement pattern and the corresponding energy gain on the supercell size. For the different supercell sizes $6 \times 6$, $12 \times 12$ and $18 \times 18$, the same stable displacement pattern can be found for a doping of four additional electrons in the supercell, see relaxed crystal structures at the top. This holds true as long as the size of the supercell can accommodate the size of the multipolaronic displacement. For all simulated supercells, the relaxation into the displacement pattern results in the essentially same binding energy of about \SI{100}{\milli\eV}, see the corresponding total energy as a function of the displacement shown below each structure.
			}
 \label{fig:SupercellSize}
\end{figure*}

\section{Metastable polaronic deformations}
\label{metastable}

The resulting size and shape of the polaronic deformation depend on the amount of doping in the investigated $18\times 18$ \mo{} supercell. While Fig.~4 of the main manuscript shows the instability of n-doped SL \mo{} towards multipolaronic deformations when introducing four additional electrons into the supercell, a doping of only three additional electrons results in metastable polaronic deformations in n-doped \mo{}, see Supplementary Fig.~\ref{fig:MetastablePolaron}. The relaxed structure in Supplementary Fig.~\ref{fig:MetastablePolaron}~(a) shows the breaking of the rotational and of most (vertical) mirror plane symmetries of the system. A doping below three additional electrons in the supercell does not result in any instability towards localized polaronic deformations.

\section{Dependence on supercell size}
\label{supercell}
Our ana\-ly\-sis shows nearly no dependence of the multipolaronic displacement pattern and the corresponding total energy gain on the size of the supercell used in the simulation. We show the relaxed crystal structures for a doping of four additional electrons on supercell sizes of $6 \times 6$, $12 \times 12$ and $18 \times 18$ in Supplementary Fig.~\ref{fig:SupercellSize} (for $18 \times 18$, see Fig.~4 as well). For each supercell with a size large enough to contain the multipolaronic distortion, the relaxation results in the same displacement pattern as shown in Fig.~4 of the main text. Below each relaxed crystal structure, the corresponding total energy as a function of the displacement can be found. The energy gained by the relaxation into the multipolaronic displacement equals a binding energy of about \SI{100}{\milli\eV} and proves independent of the supercell size as well. For supercell sizes $12 \times 12$ and $18 \times 18$, the energy barrier between the symmetric structure and the multipolaronic distortion is comparable as well, while it is shown to be smaller for $6 \times 6$.

\begin{figure}[ht]
	\centering
		\includegraphics[width=0.45\textwidth]{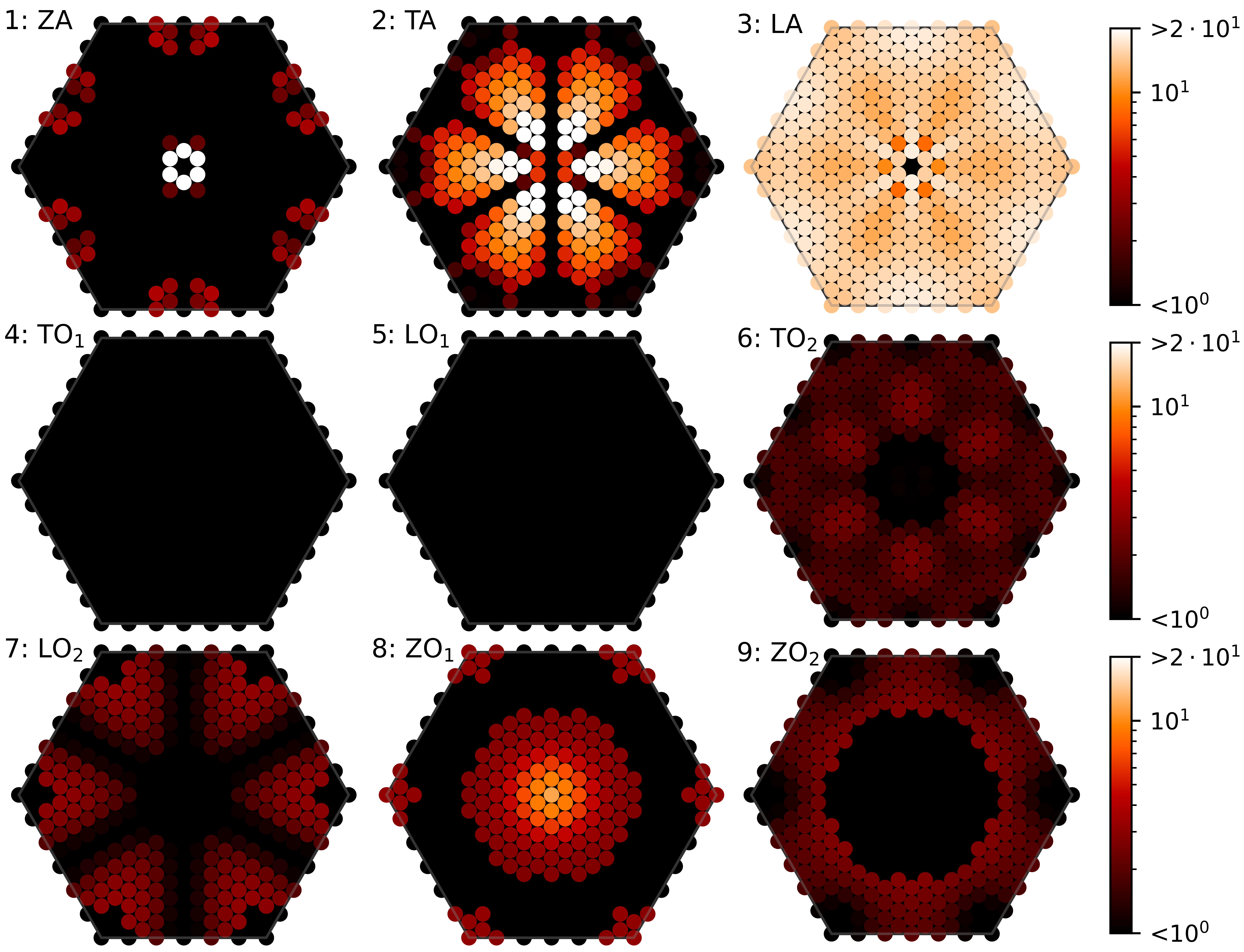}
			\caption{Decomposition of the stable polaronic distortion into normal modes.~Since the structure factor in Fig.~4~(c) of the main text shows a range of $\mathbf k$-points contributing to the localized distortion pattern, peaking around the M-point, the contributions of each normal mode to the structure factor are analyzed. The stable polaronic displacement pattern in Fig.~4~(a) of the main manuscript is decomposed into the nine normal modes, which are each labeled at the top. The bounds of the logarithmic colorbar are chosen for the best visibility of each relevant contribution. The only notable contribution between the K- and M-point stems from the LA mode (plot 3, upper right), which therefore dominates the localized polaronic displacements with four additional electrons doped.}
 \label{fig:NormalModes}
\end{figure}

\begin{figure}[tb]
	\centering
		\includegraphics[width=0.45\textwidth]{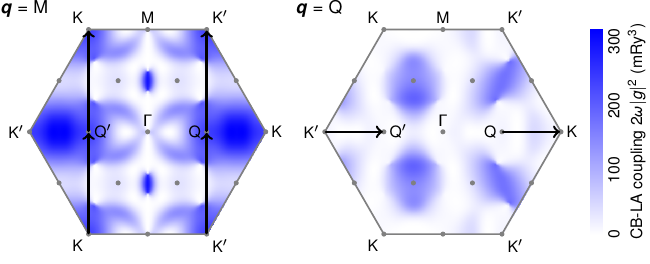}
			\caption{Visualization of the equality between the crystal momentum of M-point phonons and the distance between Q/$\mathrm Q'$ and $\mathrm K'$/K point in the Brillouin zone of \mo. The relevant high symmetry points $\Gamma$, M, K ($\mathrm K'$) and Q ($\mathrm Q'$) are labeled. It is clear that the wave vector between $\Gamma$ and M corresponds to the distance between $\mathrm Q'$ and K or between Q and $\mathrm K'$, see left plot of the figure. The squared modulus of the electron-phonon coupling between the conduction band (CB) and the LA mode as a function of $\mathbf k$ is visualized via the blue shaded background of the Brillouin zone and quantifies the scattering of electrons from $\mathbf k$ to $\mathbf k$ + $\mathbf q$. In case of $\mathbf q$ = M, the coupling strength shows significant contributions around the relevant regions. The connection between K and Q is additionally also possible via $\mathbf q$ = Q, for which the electron-phonon coupling strength shows no relevant contributions.
			}
 \label{fig:MomentumDifferences}
\end{figure}

\section{Decomposition of polaronic distortion into normal modes}
\label{decomposition}

The structure factor of the stable polaronic deformations (see Fig.~4~(c) in the main text) reveals a finite range of $\mathbf k$-points contributing to the localized distortion pattern, with a peak in reciprocal space around the M-point. To analyze the contributions stemming from each normal mode, the decomposition of the polaronic displacement pattern in Fig.~4 of the main manuscript into normal modes is shown in Supplementary Fig.~\ref{fig:NormalModes} (modes are labeled). The most notable contribution around the relevant $\mathbf k$-points stems from the LA mode (see upper right panel), which therefore dominates the polaronic displacements resulting from the doping of four additional electrons into the system. The phonon dispersion (see Fig.~4~(d)) shows a flattening of the LA mode around approximately \SI{28}{\milli\eV} between K and M, which agrees well with the experimental observation of the bosonic energy of $\Omega_0 = \SI{24 \pm 4}{\milli\eV}$.

\section{Momentum differences KQ and $\Gamma$M}
\label{MomentumDifferences}

To visualize how phonons with crystal momentum of the M-point scatter electrons between Q and K points, we sketch the Brillouin zone of \mo~with the high-symmetry points M, K ($\mathrm K'$) and Q ($\mathrm Q'$) in Supplementary Fig. \ref{fig:MomentumDifferences}. The vector connecting $\Gamma$ with M equals the distance between $\mathrm Q'$ and K or between Q and $\mathrm K'$. The blue shaded background of the Brillouin zone shows the squared modulus of the electron-phonon coupling between the conduction band and the LA mode as a function of electronic momentum $\mathbf k$. It quantifies the scattering of electrons from $\mathbf k$ to $\mathbf k$ + $\mathbf q$, where $\mathbf q$ is the phonon momentum and fixed. In the case of $\mathbf q$ = M (see left), the coupling strength in the relevant $\mathbf k$ regions (near the starting points of the arrows) is significant. There are further $\mathbf q$ values that connect the different conduction-band valleys, namely $\mathbf q$ = Q and $\mathrm Q'$ (see right), which connect Q to K and $\mathrm K'$ to $\mathrm Q'$ and vice versa. However, here the electron-phonon coupling in the relevant $\mathbf k$ regions is comparably smaller.

	\bibliography{./library}

\end{document}